\documentclass[12pt,onecolumn,notitlepage]{article}
\usepackage{amssymb}
\usepackage{amsfonts}
\usepackage{amsmath}
\usepackage{geometry}
\usepackage{setspace}
\usepackage{portland}
\usepackage{scalefnt}
\usepackage{lineno}
\usepackage{indentfirst}
\usepackage{apacite}
\usepackage{apalike}
\usepackage{fancyhdr}
\usepackage[font={small},labelfont={small}]{caption}
\usepackage{float}
\usepackage{color}
\usepackage[USenglish]{babel}
\usepackage{sectsty}

\allsectionsfont{\normalsize}
\input{tcilatex}
\begin{document}

\title{\begingroup\scalefont{.9}A Menu-Driven Software Package\\
of Bayesian Nonparametric (and Parametric) Mixed Models\\
for Regression Analysis and Density Estimation\endgroup }
\author{George Karabatsos\thanks{%
1040 W. Harrison St. (MC\ 147), Chicago, Illinois, 60304, U.S.A. E-mail:\
gkarabatsos1@gmail.com. Phone:\ (312) 413-1816. FAX:\ (312) 996-5651.} \\
University of Illinois-Chicago}
\date{June 14, 2015}
\maketitle

\begin{abstract}
Most of applied statistics involves regression analysis of data. In
practice, it is important to specify a regression model that has minimal
assumptions which are not violated by data, to ensure that statistical
inferences from the model are informative and not misleading. This paper
presents a stand-alone and menu-driven software package, Bayesian
Regression: Nonparametric and Parametric Models, constructed from MATLAB
Compiler. Currently, this package gives the user a choice from 83 Bayesian
models for data analysis. They include 47 Bayesian nonparametric (BNP)
infinite-mixture regression models; 5 BNP infinite-mixture models for
density estimation; and 31 normal random effects models (HLMs), including
normal linear models. Each of the 78 regression models handles either a
continuous, binary, or ordinal dependent variable, and can handle
multi-level (grouped) data. All 83 Bayesian models can handle the analysis
of weighted observations (e.g., for meta-analysis), and the analysis of
left-censored, right-censored, and/or interval-censored data. Each BNP
infinite-mixture model has a mixture distribution assigned one of various
BNP prior distributions, including priors defined by either the Dirichlet
process, Pitman-Yor process (including the normalized stable process), beta
(two-parameter) process, normalized inverse-Gaussian process, geometric
weights prior, dependent Dirichlet process, or the dependent
infinite-probits prior. The software user can mouse-click to select a
Bayesian model and perform data analysis via Markov chain Monte Carlo (MCMC)
sampling. After the sampling completes, the software automatically opens
text output that reports MCMC-based estimates of the model's posterior
distribution and model predictive fit to the data. Additional text and/or
graphical output can be generated by mouse-clicking other menu options. This
includes output of MCMC convergence analyses, and estimates of the model's
posterior predictive distribution, for selected functionals and values of
covariates. The software is illustrated through the BNP regression analysis
of real data.\newline
\end{abstract}

\section{\noindent \noindent Introduction}

Regression modeling is ubiquitous in empirical areas of scientific research,
because most research questions can be asked in terms of how a dependent
variable changes as a function of one or more covariates (predictors).
Applications of regression modeling involve either prediction analysis
(e.g., Dension et al., 2002\nocite{DenisonHolmesMallickSmith02}; Hastie et
al. 2009\nocite{HastieTibsFriedman09}), categorical data analysis (e.g.,
Agresti, 2002\nocite{Agresti02}), causal analysis (e.g., Imbens, 2004\nocite%
{Imbens04}; Imbens \&\ Lemieux, 2008\nocite{ImbensLemieux2008}; Stuart, 2010%
\nocite{Stuart10}), meta-analysis (e.g., Cooper, et al. 2009\nocite%
{CooperHedgesValentine09}), survival analysis of censored data (e.g., Klein
\&\ Moeschberger, 2010\nocite{KleinMoeschberger10}), spatial data analysis
(e.g., Gelfand, et al., 2010\nocite{GelfandDiggleGuttorpFuentes10}),
time-series analysis (e.g., Prado \&\ West, 2010\nocite{PradoWest2010}),
item response theory (IRT)\ analysis (e.g., van der Linden, 2015\nocite%
{vanderLinden15}), and/or other types of regression analyses.

These applications often involve either the normal random-effects
(multi-level) linear regression model (e.g., hierarchical linear model;
HLM). This general model assumes that the mean of the dependent variable
changes linearly as a function of each covariate; the distribution of the
regression errors follows a zero-mean symmetric continuous (e.g., normal)\
distribution; and the random regression coefficients are normally
distributed over pre-defined groups, according to a normal (random-effects)\
mixture distribution. Under the ordinary linear model, this mixture
distribution has variance zero. For a discrete dependent variable, all of
the previous assumptions apply for the underlying (continuous-valued)\
latent dependent variable. For example, a logit model (probit model, resp.)\
for a binary-valued (0 or 1) dependent variable implies a linear model for
the underlying latent dependent variable, with error distribution assumed to
follow a logistic distribution with mean 0 and scale 1 (normal distribution
with mean 0 and variance 1, resp.) (e.g., Dension et al., 2002\nocite%
{DenisonHolmesMallickSmith02}).

If data violate any of these linear model assumptions, then the estimates of
regression coefficient parameters can be misleading. As a result, much
research has devoted to the development of more flexible, Bayesian
nonparametric (BNP)\ regression models. Each of these models can provide a
more robust, reliable, and rich approach to statistical inference,
especially in common settings where the normal linear model assumptions are
violated. Excellent reviews of BNP\ models are given elsewhere (e.g.,
Walker, et al., 1999\nocite{WalkerDamienLaudSmith99}; Ghosh \& Ramamoorthi,
2003\nocite{GR03}; M\"{u}ller\ \&\ Quintana, 2004\nocite{MullerQuintana04};
Hjort, et al. 2010\nocite{HjortHolmesMullerWalker10}; Mitra \&\ M\"{u}ller,
2015\nocite{MitraMuller15}).

A BNP\ model is a highly-flexible model for data, defined by an infinite (or
a very large finite)\ number of parameters, with parameter space assigned a
prior distribution with large supports (M\"{u}ller \&\ Quintana, 2004\nocite%
{MullerQuintana04}). Typical BNP\ models have an infinite-dimensional,
functional parameter, such as a distribution function. According to Bayes'
theorem, a set of data updates the prior to a posterior distribution, which
conveys the plausible values of the model parameters given the data and the
chosen prior. Typically in practice, Markov chain Monte Carlo (MCMC)
sampling methods (e.g., Brooks, et al. 2011\nocite{BrooksGelmanJonesMeng11})
are used to estimate the posterior distribution (and chosen functionals)\ of
the model parameters.

Among the many BNP models that are available, the most popular models in
practice are infinite-mixture models, each having mixture distribution
assigned a (BNP)\ prior distribution on the entire space of probability
measures (distribution functions). BNP infinite-mixture models are popular
in practice, because they can provide a flexible and robust regression
analysis of data, and provide posterior-based clustering of subjects into
distinct homogeneous groups, where each subject cluster group is defined by
a common value of the (mixed)\ random model parameter(s). A standard BNP\
model is defined by the Dirichlet process (infinite-) mixture (DPM)\ model
(Lo, 1984\nocite{Lo84}), with mixture distribution assigned a Dirichlet
process (DP) (Ferguson, 1973\nocite{Ferguson73}) prior distribution on the
space of probability measures. Also, often in practice, a BNP\ model is
specified as an infinite-mixture of normal distributions. This is motivated
by the well-known fact that any smooth probability density (distribution)\
of any shape and location can be approximated arbitrarily-well by a mixture
of normal distributions, provided that the mixture has a suitable number of
mixture components, mixture weights, and component parameters (mean and
variance).

A flexible BNP\ infinite-mixture model need not be a DPM\ model, but may
instead have a mixture distribution that is assigned another BNP prior,
defined either by a more general stick-breaking process (Ishwaran \&\ James,
2001\nocite{IshwaranJames01}; Pitman, 1996\nocite{Pitman96}), such as the
Pitman-Yor (or Poisson-Dirichlet) process (Pitman, 1996\nocite{Pitman96};
Pitman \&\ Yor, 1997\nocite{PitmanYor97}), the normalized stable process
(Kingman, 1975\nocite{Kingman75}), the beta two-parameter process (Ishwaran
\&\ Zarapour, 2000\nocite{IshwaranZarepour00}); or a process with more
restrictive, geometric mixture weights (Fuentes-Garc\'{\i}a, et al. 2009,
2010\nocite{Fuentes-GarciaMenaWalker09}\nocite{Fuentes-GarciaMenaWalker10});
or defined by the normalized inverse-Gaussian process (Lijoi, et al. 2005%
\nocite{LijoiMenaPrunster05}), a general type of normalized random measure
(Regazzini et al., 2003\nocite{RegazziniLijoiPrunster03}).

A more general BNP\ infinite-mixture model can be constructed by assigning
its mixture distribution a covariate-dependent BNP\ prior. Such a BNP\
mixture model allows the entire dependent variable distribution to change
flexibly as a function of covariate(s). The Dependent Dirichlet process
(DDP; MacEachern, 1999\nocite{MacEachern99}, 2000\nocite{MacEachern00}, 2001%
\nocite{MacEachern01}) is a seminal covariate-dependent BNP\ prior. On the
other hand, the infinite-probits prior is defined by a dependent normalized
random measure, constructed by an infinite number of covariate-dependent
mixture weights, with weights specified by an ordinal probits regression
with prior distribution assigned to the regression coefficient and error
variance parameters (Karabatsos \& Walker, 2012a\nocite{KarabatsosWalker12c}%
).

The applicability of BNP\ models, for data analysis, depends on the
availability of user-friendly software. This is because BNP models typically
admit complex representations, which may not be immediately accessible to
non-experts or beginners in BNP. Currently there are two nice command-driven
R software packages for BNP\ mixture modeling. The \textbf{DPpackage} (Jara,
et al., 2011\nocite{Jara_etal11}) of R (the R Development Core Team, 2015%
\nocite{Rsoftware15}) includes many BNP models, mostly DPM\ models, that
provide either flexible regression or density estimation for data analysis.
The package also provides BNP\ models having parameters assigned a flexible
mixture of finite P\'{o}lya Trees BNP\ prior (Hanson, 2006\nocite{Hanson06}%
). The \textbf{bspmma} R package (Burr, 2012\nocite{Burr12}) provides DPM\
normal-mixture models for meta-analysis.

The existing packages for BNP\ modeling, while impressive, still suggest
room for improvements, as summarized by the following points.

\begin{enumerate}
\item While the existing BNP\ packages provide many DPM\ models, they do not
provide a BNP\ infinite-mixture model with mixture distribution assigned any
one of the other important BNP\ priors mentioned earlier. Priors include
those defined by the Pitman-Yor, normalized stable, beta, normalized
inverse-Gaussian process; or defined by a geometric weights or
infinite-probits prior. As an exception, the \textbf{DPpackage} provides a
Pitman-Yor process mixture of regressions model for interval-censored data
(Jara, et al. 2010\nocite{JaraEtAl10}).

\item The \textbf{bspmma} R package (Burr, 2012\nocite{Burr12}), for
meta-analysis, is limited to DPM\ models that do not incorporate covariate
dependence (Burr \&\ Doss, 2005\nocite{Burr_Doss05}).

\item The \textbf{DPpackage} handles interval-censored data, but does not
handle left- or right-censored data.

\item While both BNP\ packages use MCMC\ sampling algorithms to estimate the
posterior distribution of the user-chosen model, each package does not
provide a MCMC\ convergence analysis (e.g., Flegal \&\ Jones, 2011\nocite%
{FlegalJones11}). A BNP\ package that provides its own menu options for
MCMC\ convergence analysis would be, for the user, faster and more
convenient, and would not require learning a new package (e.g., \textbf{CODA}
R package; Plummer et al., 2006\nocite{PlummerBestCowlesVines06}) to conduct
MCMC\ convergence analyses.

\item Both BNP\ packages do not provide many options to investigate how the
posterior predictive distribution (and chosen functionals)\ of the dependent
variable, varies as a function of one or more covariates.

\item Generally speaking, command-driven software can be unfriendly,
confusing, and time-consuming to beginners and to experts.
\end{enumerate}

In this paper, we introduce a stand-alone and user-friendly software package
for BNP\ modeling, which the author constructed using MATLAB Compiler
(Natick, MA). This package, named:\ \textbf{Bayesian Regression:\
Nonparametric and Parametric Models}, provides BNP\ data analysis in a fully
menu-driven software environment that resembles SPSS (I.B.M., 2015\nocite%
{IBMcorp15}).

\noindent The software allows the user to mouse-click menu options:

\begin{enumerate}
\item To inspect, describe, and explore the variables of the data set, via
basic descriptive statistics (e.g., means, standard deviations,
quantiles/percentiles) and graphs (e.g., scatter plots, box plots, normal
Q-Q plots, kernel density plots, etc.);

\item To pre-process the data of the dependent variable and/or the
covariate(s) before including the variable(s) into the BNP\ regression model
for data analysis. Examples of data pre-processing includes constructing new
dummy indicator (0 or 1) variables and/or two-way interaction variables from
the covariates (variables), along with other options to transform variables;
and performing a nearest-neighbor hot-deck imputation (Andridge \&\ Little,
2010\nocite{AndridgeLittle10}) of missing data values in the variables
(e.g., covariate(s)).

\item To use list and input dialogs to select, in the following order: the
Bayesian model for data analysis; the dependent variable; covariate(s) (if a
regression model was selected); parameters of the prior distribution of the
model; the (level-2 and possibly level-3)\ grouping variables (for a
multilevel model, if selected); the observation weights variable (if
necessary; e.g., to set up a meta-analysis); and the variables describing
the nature of the censored dependent variable observations (if necessary;
e.g., to set up a survival analysis). The observations can either be
left-censored, right-censored, interval-censored, or uncensored. Also, if so
desired, the user can easily use point-and-click to quickly highlight and
select a large list of covariates for the model, whereas command-driven
software requires the user to carefully type (or copy and paste) and
correctly-verify the long list of the covariates.
\end{enumerate}

\noindent After the user make these selections, \noindent the Bayesian
Regression software immediately presents a graphic of the user-chosen
Bayesian model in the middle of the computer screen, along with all of the
variables that were selected for this model (e.g., dependent variables,
covariate(s); see \#3\ above). The explicit presentation of the model is
important because BNP\ models typically admit complex representations. In
contrast, the command-driven packages do not provide immediate on-screen
presentations of the BNP\ model selected by the user.

Then the software user can click a button to run the MCMC\ sampling
algorithm for the menu-selected Bayesian model. The user clicks this button
after entering a number of MCMC\ sampling iterations. Immediately after all
the MCMC\ sampling iterations have completed, the software automatically
opens a text output file that summarizes the basic results of the data
analysis (derived from the generated MCMC\ samples). Results include
point-estimates of the (marginal)\ posterior distributions of the model's
parameters, and summaries of the model's predictive fit to the data. Then,
the user can click other menu options to produce graphical output of the
results. They include density plots, box plots, scatter plots, trace plots,
and various plots of (marginal)\ posterior distributions of model parameters
and fit statistics. For each available BNP\ infinite-mixture model, the
software implements standard slice sampling MCMC\ methods (Kalli,\ et al.
2011\nocite{KalliGriffinWalker11}) that are suitable for making inferences
of the posterior distribution (and chosen functionals)\ of model parameters.

Next, after a few mouse-clicks of appropriate menu options, the user can
perform a detailed MCMC\ convergence analysis. This analysis evaluates
whether a sufficiently-large number of MCMC\ samples (sampling iterations of
the MCMC\ algorithm) has been generated, in order to warrant the conclusion
that these samples have converged to samples from the posterior distribution
(and chosen functionals)\ of the model parameters. More details about how to
use the software to perform MCMC\ convergence analysis is provided in
Sections 3.1 and 5.2.

The software also provides menu options to investigate how the posterior
predictive distribution (and functionals)\ of the dependent variable changes
as a function of covariates. Functionals of the posterior predictive
distribution include: the mean, median, and quantiles to provide a quantile
regression analysis; the variance functional to prove a variance regression
analysis; the probability density function (p.d.f.) and the cumulative
distribution function (c.d.f.) to provide a density regression analysis; and
the survival function, hazard function, and the cumulative hazard function,
for survival analysis. The software also provides posterior predictive
inferences for BNP\ infinite-mixture models that do not incorporate
covariates and only focus on density estimation.

Currently, the \textbf{Bayesian Regression} software provides the user a
choice from 83 Bayesian models for data analysis. Models include 47 BNP\
infinite-mixture regression models, 31 normal linear models for comparative
purposes, and 5 BNP\ infinite normal mixture models for density estimation.
Most of the infinite-mixture models are defined by normal mixtures.

The 47 BNP\ infinite-mixture regression models can each handle a dependent
variable that is either continuous-valued, binary-valued (0 or 1), or
ordinal valued ($c=0,1,\ldots ,m$), using either a probit or logit version
of this model for a discrete dependent variable; with mixture distribution
assigned a prior distribution defined either by the Dirichlet process,
Pitman-Yor process (including the normalized stable process), beta
(2-parameter)\ process, geometric weights prior, normalized inverse-Gaussian
process, or an infinite-probits regression prior; and with mixing done on
either the intercept parameter, or on the intercept and slope coefficient
parameters, and possibly on the error variance parameter. Specifically, the
regression models with mixture distribution assigned a Dirichlet process
prior are equivalent to ANOVA/linear DDP models, defined by an
infinite-mixture of normal distributions, with a covariate-dependent mixture
distribution defined by independent weights (De Iorio, et al. 2004\nocite%
{DeIorioMullerRosnerMacEachern04}; M\"{u}ller et al., 2005\nocite%
{MullerRosnerDeIorioMacEachern05}). Similarly, the models with mixture
distribution, instead, assigned a different BNP\ prior distribution
(process)\ mentioned above, implies a covariate-dependent version of that
process. See Section 3.1 for more details. Also, some of the
infinite-mixture regression models, with covariate-dependent mixture
distribution assigned a infinite-probits prior, have spike-and-slab priors
assigned to the coefficients of this BNP\ prior, based on stochastic search
variable selection (SSVS) (George \& McCulloch, 1993\nocite%
{GeorgeMcCulloch93}, 1997\nocite{GeorgeMcCulloch97}). In addition, the 5
BNP\ infinite normal mixture models, for density estimation, include those
with mixture distribution assigned a BNP\ prior distribution that is defined
by either one of the 5 BNP\ process mentioned above (excluding
infinite-probits).

Finally, the 31 Bayesian linear models of the Bayesian regression software
include ordinary linear models, 2-level, and 3-level normal random-effects
(or HLM)\ models, for a continuous dependent variable; probit and logit
versions of these linear models for\ either a binary (0 or 1) or ordinal ($%
c=0,1,\ldots ,m$) dependent variable; and with mixture distribution
specified for the intercept parameter, or for the intercept and slope
coefficient parameters.

The outline for the rest of the paper is as follows. Section 2 reviews the
Bayesian inference framework. Appendix A\ reviews the basic probability
theory notation and concepts that we use. In Section 3, we define two key
BNP\ infinite-mixture regression models, each with mixture distribution
assigned a BNP\ prior distribution on the space of probability measures. The
other 50 BNP infinite-mixture models of the \textbf{Bayesian regression}
software are extensions of these two key models, and in that section we give
an overview of the various BNP\ priors mentioned earlier. In that section we
also describe the Bayesian normal linear model, and a Bayesian normal
random-effects linear model (HLM). Section 4 gives step-by-step software
instructions on how to perform data analysis using a menu-chosen, Bayesian
model. Section 5 illustrates the \textbf{Bayesian regression} software
through the analysis of a real data set, using each of the two key BNP\
models, and a Bayesian linear model. Appendix B provides a list of exercises
that the software user can work through in order to practice BNP\ modeling
on several example data sets, available from the software. These
data-analysis exercises address applied problems in prediction analysis,
categorical data analysis, causal analysis, meta-analysis, survival analysis
of censored data, spatial data analysis, time-series analysis, and item
response theory analysis. Section 6 ends with conclusions.

\section{Overview of Bayesian Inference}

In a given research setting where it is of interest to apply a regression
data analysis, a sample data set is of the form $\mathcal{D}_{n}=\{(y_{i},%
\mathbf{x}_{i})\}_{i=1}^{n}$. Here, $n$ is the sample size of the
observations, respectively indexed by $i=1,\ldots ,n$, where $y_{i}$ is the $%
i$th observation of the dependent variable $Y_{i}$, corresponding to an
observed vector of $p$ observed covariates\footnote{%
For each Bayesian model for density estimation, from the software, we assume 
$\mathbf{x}=1$.} $\mathbf{x}_{i}=(1,x_{i1},\ldots ,x_{ip})^{\intercal }$. A
constant (1)\ term is included in $\mathbf{x}$ for future notational
convenience.

A regression model assumes a specific form for the probability density (or
p.m.f.)\ function $f(y\,|\,\mathbf{x};\boldsymbol{\zeta })$, conditionally
on covariates $\mathbf{x}$ and model parameters denoted by a vector, $%
\boldsymbol{\zeta }\in \Omega _{\boldsymbol{\zeta }}$, where $\Omega _{%
\boldsymbol{\zeta }}=\{\boldsymbol{\zeta }\}$ is the parameter space. For
any given model parameter value $\boldsymbol{\zeta }\in \Omega _{\boldsymbol{%
\zeta }}$, the density $f(y_{i}\,|\,\mathbf{x}_{i};\boldsymbol{\zeta })$ is
the likelihood of $y_{i}$ given $\mathbf{x}_{i}$, and $L(\mathcal{D}_{n}\,;\,%
\boldsymbol{\zeta })=\tprod\nolimits_{i=1}^{n}f(y_{i}\,|\,\mathbf{x}_{i};%
\boldsymbol{\zeta })$ is the likelihood of the full data set $\mathcal{D}%
_{n} $ under the model. A Bayesian regression model is completed by the
specification of a prior distribution (c.d.f.) $\Pi (\boldsymbol{\zeta })$
over the parameter space $\Omega _{\boldsymbol{\zeta }}$, and $\pi (%
\boldsymbol{\zeta })$ gives the corresponding probability density of a given
parameter $\boldsymbol{\zeta }\in \Omega _{\boldsymbol{\zeta }}$.

According to Bayes' theorem, after observing the data $\mathcal{D}%
_{n}=\{(y_{i},\mathbf{x}_{i})\}_{i=1}^{n}$, the plausible values of the
model parameter $\boldsymbol{\zeta }$ is given by the posterior
distribution. This distribution defines the posterior probability density of
a given parameter $\boldsymbol{\zeta }\in \Omega _{\boldsymbol{\zeta }}$ by:%
\begin{equation}
\pi (\boldsymbol{\zeta }\,|\,\mathcal{D}_{n})=\dfrac{\tprod%
\nolimits_{i=1}^{n}f(y_{i}\,|\,\mathbf{x}_{i};\boldsymbol{\zeta })\mathrm{d}%
\Pi (\boldsymbol{\zeta })}{\dint\nolimits_{\Omega _{\boldsymbol{\zeta }%
}}\tprod\nolimits_{i=1}^{n}f(y_{i}\,|\,\mathbf{x}_{i};\boldsymbol{\zeta })%
\mathrm{d}\Pi (\boldsymbol{\zeta })}.  \label{Posterior}
\end{equation}%
Conditionally on a chosen value of the covariates $\mathbf{x}%
=(1,x_{1},\ldots ,x_{p})^{\intercal }$, the posterior predictive density of
a future observation $y_{n+1}$, and the corresponding posterior predictive
c.d.f. ($F(y\,|\,\mathbf{x})$), mean (expectation, $\mathbb{E}$), variance ($%
\mathbb{V}$), median, $u$th quantile ($Q(u\,|\,\mathbf{x})$, for some chosen 
$u\in \lbrack 0,1]$, with $Q(.5\,|\,\mathbf{x})$ the conditional median),
survival function ($S$), hazard function ($H$), and cumulative hazard
function ($\Lambda $), is given respectively by: 
\begin{subequations}
\label{pp all}
\begin{eqnarray}
f_{n}(y\,|\,\mathbf{x}) &=&\dint f(y\,|\,\mathbf{x};\boldsymbol{\zeta })%
\mathrm{d}\Pi (\boldsymbol{\zeta }\,|\,\mathcal{D}_{n}),  \label{pp pdf} \\
F_{n}(y\,|\,\mathbf{x}) &=&\dint_{Y\leq y}f(y\,|\,\mathbf{x};\boldsymbol{%
\zeta })\mathrm{d}\Pi (\boldsymbol{\zeta }\,|\,\mathcal{D}_{n}),
\label{pp cdf} \\
\mathbb{E}_{n}(Y\,|\,\mathbf{x}) &=&\dint y\mathrm{d}F_{n}(y\,|\,\mathbf{x}),
\label{pp E} \\
\mathbb{V}_{n}(Y\,|\,\mathbf{x}) &=&\dint \{y-\mathbb{E}_{n}(Y\,|\,\mathbf{x}%
)\}^{2}\mathrm{d}F_{n}(y\,|\,\mathbf{x}),  \label{pp Var} \\
Q_{n}(u\,|\,\mathbf{x}) &=&F_{n}^{-1}(u\,|\,\mathbf{x}),  \label{pp q} \\
S_{n}(y\,|\,\mathbf{x}) &=&1-F_{n}(y\,|\,\mathbf{x}),  \label{pp surv} \\
H_{n}(y\,|\,\mathbf{x}) &=&f_{n}(y\,|\,\mathbf{x})/\{1-F_{n}(y\,|\,\mathbf{x}%
)\},  \label{pp haz} \\
\Lambda _{n}(y\,|\,\mathbf{x}) &=&-\log \{1-F_{n}(y\,|\,\mathbf{x})\}.
\label{pp cumhaz}
\end{eqnarray}%
Depending on the choice of posterior predictive functional from (\ref{pp all}%
), a Bayesian regression analysis can provide inferences in terms of how the
mean (\ref{pp E}), variance (\ref{pp Var}), quantile (\ref{pp q}) (for a
given choice $u\in \lbrack 0,1]$), p.d.f. (\ref{pp pdf}), c.d.f. (\ref{pp
cdf}), survival function (\ref{pp surv}), hazard function (\ref{pp haz}), or
cumulative hazard function (\ref{pp cumhaz}), of the dependent variable $Y$,
varies as a function of the covariates $\mathbf{x}$. While the mean
functional $\mathbb{E}_{n}(Y\,|\,\mathbf{x})$ is conventional for applied
regression, the choice of functional $\mathbb{V}_{n}(Y\,|\,\mathbf{x})$
pertains to variance regression; the choice of function $Q_{n}(u\,|\,\mathbf{%
x})$ pertains to quantile regression; the choice of p.d.f. $f_{n}(y\,|\,%
\mathbf{x})$ or c.d.f. $F_{n}(y\,|\,\mathbf{x})$ pertains to Bayesian
density (distribution)\ regression; and the choice of survival $S_{n}(y\,|\,%
\mathbf{x})$ or a hazard function ($H_{n}(y\,|\,\mathbf{x})$ or $\Lambda
_{n}(y\,|\,\mathbf{x})$) pertains to survival analysis.

In practice, the predictions of the dependent variable $Y$ (for a chosen
functional from (\ref{pp all})), can be easily viewed (in a graph or table)\
as a function of a subset of only one or two covariates. Therefore, for
practice we need to consider predictive methods that involve such a small
subset of covariates. To this end, let $\mathbf{x}_{\mathcal{S}}$ be a focal
subset of the covariates $(x_{1},\ldots ,x_{p})$, with $\mathbf{x}_{\mathcal{%
S}}$ also including the constant (1) term. Also,\ let $\mathbf{x}_{\mathcal{C%
}}$ be the non-focal, complement set of $q$ (unselected) covariates. Then $%
\mathbf{x}_{\mathcal{S}}\cap \mathbf{x}_{\mathcal{C}}\neq \emptyset $ and $%
\mathbf{x=x}_{\mathcal{S}}\cup \mathbf{x}_{\mathcal{C}}$.

It is possible to study how the predictions of a dependent variable $Y$
varies as a function of the focal covariates $\mathbf{x}_{\mathcal{S}}$,
using one of four automatic methods. The first two methods are conventional.
They include the \textit{grand-mean centering method}, which assumes that
the non-focal covariates $\mathbf{x}_{\mathcal{C}}$ is defined by the mean
in the data $\mathcal{D}_{n}$, with $\mathbf{x}_{\mathcal{C}}:=\tfrac{1}{n}%
\tsum\nolimits_{i=1}^{n}\mathbf{x}_{\mathcal{C}i}$; and the \textit{%
zero-centering method}, which assumes that the non-focal covariates are
given by $\mathbf{x}_{\mathcal{C}}:=\mathbf{0}_{q}$ where $\mathbf{0}_{q}$
is a vector of $q$ zeros. Both methods coincide if the observed covariates $%
\{\mathbf{x}_{i}=(1,x_{i1},\ldots ,x_{ip})^{\intercal }\}_{i=1}^{n}$ in the
data $\mathcal{D}_{n}$ have average $(1,0,\ldots ,0)^{\intercal }$. This is
the case if the covariate data $\{x_{ik}\}_{i=1}^{n}$ have already been
centered to have mean zero, for $k=1,\ldots ,p$.

The partial dependence method (Friedman, 2001, Section 8.2\nocite{Friedman01}%
) is the third method for studying how the predictions of a dependent
variable $Y$ varies as a function of the focal covariates $\mathbf{x}_{%
\mathcal{S}}$. In this method, the predictions of $Y$, conditionally on each
value of the focal covariates $\mathbf{x}_{\mathcal{S}}$, are averaged over
data ($\mathcal{D}_{n}$) observations $\{\mathbf{x}_{\mathcal{C}%
i}\}_{i=1}^{n}$ (and effects) of the non-focal covariates $\mathbf{x}_{%
\mathcal{C}}$. Specifically, in terms of the posterior predictive
functionals (\ref{pp all}), the averaged prediction of $Y$, conditionally on
a value of the covariates $\mathbf{x}_{\mathcal{S}}$, is given respectively
by: 
\end{subequations}
\begin{subequations}
\label{pd all}
\begin{eqnarray}
f_{n}(y\,|\,\mathbf{x}_{S}) &=&\tfrac{1}{n}\tsum\nolimits_{i=1}^{n}f_{n}(y%
\,|\,\mathbf{x}_{S},\mathbf{x}_{\mathcal{C}i}),  \label{pd pdf} \\
F_{n}(y\,|\,\mathbf{x}_{S}) &=&\tfrac{1}{n}\tsum\nolimits_{i=1}^{n}F_{n}(y%
\,|\,\mathbf{x}_{S},\mathbf{x}_{\mathcal{C}i}),  \label{pd cdf} \\
\mathbb{E}_{n}(Y\,|\,\mathbf{x}_{S}) &=&\tfrac{1}{n}\tsum\nolimits_{i=1}^{n}%
\mathbb{E}_{n}(Y\,|\,\mathbf{x}_{S},\mathbf{x}_{\mathcal{C}i}),  \label{pd E}
\\
\mathbb{V}_{n}(Y\,|\,\mathbf{x}_{S}) &=&\tfrac{1}{n}\tsum\nolimits_{i=1}^{n}%
\mathbb{E}_{n}(Y\,|\,\mathbf{x}_{S},\mathbf{x}_{\mathcal{C}i}),  \label{pd V}
\\
Q_{n}(u\,|\,\mathbf{x}_{S}) &=&\tfrac{1}{n}\tsum%
\nolimits_{i=1}^{n}F_{n}^{-1}(u\,|\,\mathbf{x}_{S},\mathbf{x}_{\mathcal{C}%
i}),  \label{pd Q} \\
S_{n}(y\,|\,\mathbf{x}_{S}) &=&\tfrac{1}{n}\tsum\nolimits_{i=1}^{n}%
\{1-F_{n}(y\,|\,\mathbf{x}_{S},\mathbf{x}_{\mathcal{C}i})\},  \label{pd Surv}
\\
H_{n}(y\,|\,\mathbf{x}_{S}) &=&\tfrac{1}{n}\tsum\nolimits_{i=1}^{n}H_{n}(y%
\,|\,\mathbf{x}_{\mathcal{S}},\mathbf{x}_{\mathcal{C}i}),  \label{pd Haz} \\
\Lambda _{n}(y\,|\,\mathbf{x}_{S}) &=&\tfrac{1}{n}\tsum\nolimits_{i=1}^{n}-%
\log \{1-F_{n}(y\,|\,\mathbf{x}_{S},\mathbf{x}_{\mathcal{C}i})\}.
\label{pd CumHaz}
\end{eqnarray}%
The equations above give, respectively, the (partial dependence)\ posterior
predictive density, c.d.f., mean, variance, quantile (at $u\in \lbrack 0,1]$%
), survival function, hazard function, and cumulative hazard function, of $Y$%
, conditionally on a value $\mathbf{x}_{S}$ of the focal covariates. As a
side note pertaining to causal analysis, suppose that the focal covariates
include a covariate, denoted $T$, along with a constant (1)\ term, so that $%
\mathbf{x}_{S}=(1,t)$. Also suppose that the covariate $T$ is a
binary-valued (0,1) indicator of treatment receipt, versus non-treatment
receipt. Then the estimate of a chosen (partial-dependence) posterior
predictive functional of $Y$\ under treatment ($T=1$) from (\ref{pd all}),
minus that posterior predictive functional under control ($T=0$), provides
an estimate of the causal average treatment effect (CATE). This is true
provided that the assumptions of unconfoundedness and overlap hold (Imbens,
2004\nocite{Imbens04}).

The partial-dependence method can be computationally-demanding, as a
function of sample size ($n$), the dimensionality of $\mathbf{x}_{S}$, the
number of $\mathbf{x}_{S}$ values considered when investigating how $Y$\
varies as a function of $\mathbf{x}_{S}$, and the number of MCMC\ sampling
iterations performed for the estimation of the posterior distribution
(density (\ref{Posterior})) of the model parameters. In contrast, the 
\textit{clustered partial dependence method}, the fourth method, is less
computationally-demanding. This method is based on forming $K$-means cluster
centroids, $\{\mathbf{x}_{\mathcal{C}t}\}_{t=1}^{K}$, of the data
observations $\{\mathbf{x}_{\mathcal{C}i}\}_{i=1}^{n}$ of the non-focal
covariates $\mathbf{x}_{\mathcal{C}}$, with $K=\sqrt{n/2}$ clusters as a
rule-of-thumb. Then the posterior predictions of $Y$, conditionally on
chosen value of the covariate subset $\mathbf{x}_{\mathcal{S}}$, is given by
any one of the chosen posterior functionals (\ref{pd all}) of interest,
after replacing $\tfrac{1}{n}\tsum\nolimits_{i=1}^{n}$ with $\tfrac{1}{K}%
\tsum\nolimits_{t=1}^{K}$, and $\mathbf{x}_{\mathcal{C}i}$ with $\mathbf{x}_{%
\mathcal{C}t}$.

The predictive fit of a Bayesian regression model, to a set of data, $%
\mathcal{D}_{n}=\{(y_{i},\mathbf{x}_{i})\}_{i=1}^{n}$, can be assessed on
the basis of the posterior predictive expectation (\ref{pp E}) and variance (%
\ref{pp Var}). First, the standardized residual fit statistics of the model
are defined by: 
\end{subequations}
\begin{equation}
r_{i}=\dfrac{y_{i}-\mathbb{E}_{n}(Y\,|\,\mathbf{x}_{i})}{\sqrt{\mathbb{V}%
_{n}(Y\,|\,\mathbf{x}_{i})}},\text{ \ \ }i=1,\ldots ,n.  \label{pp residual}
\end{equation}%
An observation $y_{i}$ can be judged as an outlier under the model, when its
absolute standardized residual $|\,r_{i}\,|\,$ exceeds 2 or 3. The
proportion of variance explained in the dependent variable $Y$, by a
Bayesian model, is measured by the R-squared statistic:%
\begin{equation}
R^{2}=1-\left( \dfrac{\tsum\nolimits_{i=1}^{n}(y_{i}-\mathbb{E}_{n}[Y\,|\,%
\mathbf{x}_{i}])^{2}}{\tsum\nolimits_{i=1}^{n}\left\{ y_{i}-\left( \frac{1}{n%
}\tsum\nolimits_{i=1}^{n}y_{i}\right) \right\} ^{2}}\right) .
\label{R squared}
\end{equation}%
Also, suppose that it is of interest to compare $M$ regression models, in
terms of predictive fit to the given data set $\mathcal{D}_{n}$. Models are
indexed by $\underline{m}=1,\ldots ,M$, respectively. For each model $%
\underline{m}$, a global measure of predictive fit is given by the
mean-squared predictive error criterion:%
\begin{equation}
D(\underline{m})=\tsum\nolimits_{i=1}^{n}\{y_{i}-\mathbb{E}_{n}(Y\,|\,%
\mathbf{x}_{i}\mathbf{,}\underline{m})\}^{2}+\tsum\nolimits_{i=1}^{n}\mathbb{%
V}_{n}(Y\,|\,\mathbf{x}_{i}\mathbf{,}\underline{m})  \label{Dm criterion}
\end{equation}%
(Laud \&\ Ibrahim, 1995\nocite{LaudIbrahim1995}; Gelfand \&\ Ghosh, 1998%
\nocite{GelfandGhosh98}). The first term in (\ref{Dm criterion}) measures
model goodness-of-fit to the data $\mathcal{D}_{n}$, and the second term is
a model complexity penalty. Among a set of $M$ regression models compared,
the model with the best predictive fit for the data $\mathcal{D}_{n}$ is
identified as the one that has the smallest value of $D(\underline{m})$.

\subsection{MCMC\ Methods}

In practice, a typical Bayesian model does not admit a closed-form solution
for its posterior distribution (density function of the form (\ref{Posterior}%
)). However, the posterior distribution, along with any function of the
posterior distribution of interest, can be estimated through the use of
Monte Carlo methods. In practice, they usually involve Markov Chain Monte
Carlo (MCMC) methods (e.g., Brooks et al., 2011\nocite%
{BrooksGelmanJonesMeng11}). Such a method aims to construct a discrete-time
Harris ergodic Markov chain $\{\boldsymbol{\zeta }^{(s)}\}_{s=1}^{S}$ with
stationary (posterior)\ distribution $\Pi (\boldsymbol{\zeta }\,|\,\mathcal{D%
}_{n})$, and ergodicity is ensured by a proper (integrable)\ prior density
function $\pi (\boldsymbol{\zeta })$ (Robert \&\ Casella, 2004, Section
10.4.3\nocite{RobertCasella04}). A realization $\boldsymbol{\zeta }^{(s)}$
from the Markov chain can be generated by first specifying partitions
(blocks) $\boldsymbol{\zeta }_{b}$ ($b=1,\ldots ,B$) of the model's
parameter $\boldsymbol{\zeta }$, and then simulating a sample from each of
the full conditional posterior distributions $\Pi (\boldsymbol{\zeta }%
_{b}\,|\,\mathcal{D}_{n},\boldsymbol{\zeta }_{c},c\neq b)$, in turn for $%
b=1,\ldots ,B$. Then, as $S\rightarrow \infty $, the Markov (MCMC)\ chain $\{%
\boldsymbol{\zeta }^{(s)}\}_{s=1}^{S}$ converges to samples from the
posterior distribution $\Pi (\boldsymbol{\zeta }\,|\,\mathcal{D}_{n})$.
Therefore, in practice, the goal is to construct a MCMC\ chain (samples) $\{%
\boldsymbol{\zeta }^{(s)}\}_{s=1}^{S}$\ for a sufficiently-large finite $S$.

MCMC\ convergence analyses can be performed in order to check whether a
sufficiently-large number ($S$) of sampling iterations has been run, to
warrant the conclusion that the resulting samples ($\{\boldsymbol{\zeta }%
^{(s)}\}_{s=1}^{S}$) have converged (practically)\ to samples from the
model's posterior distribution. Such an analysis may focus only on the model
parameters of interest for data analysis, if so desired. MCMC convergence
can be investigated in two steps (Geyer, 2011\nocite{Geyer11}). One step is
to inspect, for each of these model parameters, the univariate trace plot of
parameter samples over the MCMC sampling iterations. This is done to
evaluate MCMC\ mixing, i.e., the degree to which MCMC\ parameter samples
explores the parameter's support in the model's posterior distribution. Good
mixing is suggested by a univariate trace plot that appears stable and
"hairy" over MCMC iterations\footnote{%
The CUSUM\ statistic, which ranges between 0 and 1, is a measure of the
"hairiness" of a univariate trace plot of a model parameter (Brooks, 1998%
\nocite{Brooks98}). A CUSUM\ value of .5 indicates optimal MCMC\ mixing.}.
The other step is to conduct, for each model parameter of interest, a batch
means (or subsampling)\ analysis of the MCMC\ samples, in order to calculate
95\%\ Monte Carlo Confidence Intervals (95\%\ MCCIs) of posterior
point-estimates of interest (such as marginal posterior means, variances,
quantiles, etc., of the parameter) (Flegal \&\ Jones, 2011\nocite%
{FlegalJones11}). For a given (marginal) posterior point-estimate of a
parameter, the 95\%\ MCCI half-width size reflects the imprecision of the
estimate due to Monte Carlo sampling error. The half-width becomes smaller
as number of MCMC\ sampling iterations grows. In all, MCMC\ convergence is
confirmed by adequate MCMC\ mixing and practically-small 95\%\ MCCIs
half-widths (e.g., .10 or .01)\ for the (marginal)\ posterior
point-estimates of parameters (and chosen functionals)\ of interest. If
adequate convergence cannot be confirmed after a MCMC\ sampling run, then
additional MCMC\ sampling iterations can be run until convergence is
obtained for the (updated) total set of MCMC\ samples.

For each BNP\ infinite-mixture model, the \textbf{Bayesian Regression}
software estimates the posterior distribution (and functionals)\ of the
model on the basis of a general slice-sampling MCMC\ method, which can
handle the infinite-dimensional model parameters (Kalli,\ et al., 2011\nocite%
{KalliGriffinWalker11}). This slice-sampling method does so by introducing
latent variables into the likelihood function of the infinite-mixture model,
such that, conditionally on these variables, the model is finite-dimensional
and hence tractable by a computer. Marginalizing over the distribution of
these latent variables recovers the original likelihood function of the
infinite-mixture model.

We now describe the MCMC\ sampling methods that the software uses to sample
from the full conditional posterior distributions of the parameters, for
each model that the software provides. For each DPM\ model, the full
conditional posterior distribution of the unknown precision parameter ($%
\alpha $) is sampled from a beta mixture of two gamma distributions (Escobar
\&\ West, 1995\nocite{EscobarWest95}). For each BNP\ infinite-mixture model
based on a DP, Pitman-Yor process (including the the normalized stable
process), or beta process prior, the full conditional posterior distribution
of the mixture weight parameters are sampled from appropriate beta
distributions (Kalli,\ et al., 2011\nocite{KalliGriffinWalker11}). Also, for
the parameters of each of the 31 linear models, and for the linear
parameters of each of the BNP\ infinite-mixture models, the software
implements (direct) MCMC\ Gibbs sampling of standard full conditional
posterior distributions, derived from the standard theories of the Bayesian
normal linear, probit, and logit models, as appropriate (Evans, 1965\nocite%
{Evans65}; Lindley \&\ Smith, 1972\nocite{LindleySmith72}; Gilks et al., 1993%
\nocite{Gilks_etal_1993}; Albert \&\ Chib, 1993\nocite{AlbertChib93};
Bernardo \&\ Smith, 1994\nocite{BernardoSmith94}; Denison et al., 2002\nocite%
{DenisonHolmesMallickSmith02}; Cepeda \&\ Gamerman, 2001\nocite%
{CepedaGamerman01}; O'Hagan \&\ Forster, 2004\nocite{OhaganForster04};
Holmes \&\ Held, 2006\nocite{HolmesHeld06}; George \&\ McCulloch, 1997\nocite%
{GeorgeMcCulloch97}; e.g., see Karabatsos \&\ Walker, 2012a,b\nocite%
{KarabatsosWalker12a}\nocite{KarabatsosWalker12c}). When the full
conditional posterior distribution of the model parameter(s)\ is
non-standard, the software implements a rejection sampling algorithm.
Specifically, it implements an adaptive random-walk Metropolis-Hastings
(ARWMH) algorithm (Atchad\'{e} \&\ Rosenthal, 2005\nocite{AtchadeRosenthal05}%
) with normal proposal distribution, to sample from the full conditional
posterior distribution(s) of the mixture weight parameter of a BNP\
geometric weights infinite-mixture model; the mixture weight parameter of a
BNP\ normalized inverse-Gaussian process mixture model, using the equivalent
stick-breaking representation of this process (Favaro, et al. 2012\nocite%
{FavaroLijoiPrunster12}). Also, for BNP\ infinite-mixture models and normal
random-effects models that assign a uniform prior distribution to the
variance parameter for random intercepts (or means), the software implements
the slice sampling (rejection)\ algorithm with stepping-out procedure (Neal,
2003\nocite{Neal03b}), in order to sample from the full conditional
posterior distribution of this parameter. Finally, for computational speed
considerations, we use the ARWMH algorithm instead of Gibbs sampling, in
order to sample from the full conditional posterior distributions for the
random coefficient parameters (the intercepts $u_{0h}$; and possibly the $%
u_{kh},k=0,1,\ldots ,p$, as appropriate, for groups $h=1,\ldots ,H$)\ in a
normal random-effects (or random intercepts)\ HLM; and for the random
coefficients ($\boldsymbol{\beta }_{j}$) or random intercept parameters ($%
\beta _{0j}$) in a BNP\ infinite-mixture regression model, as appropriate
(Karabatsos \&\ Walker, 2012a,b\nocite{KarabatsosWalker12c}\nocite%
{KarabatsosWalker12a}).

The given data set ($\mathcal{D}_{n}$) may consist of censored dependent
variable observations (either left-, right-, and/or interval-censored). If
the software user indicates the censored dependent variable observations
(see Section 4.2, Step\ 6), then the software adds a Gibbs sampling step to
the MCMC\ algorithm, that draws from the full-conditional posterior
predictive distributions (density function (\ref{pp pdf})) to provide
multiple MCMC-based imputations of these missing censored observations
(Gelfand, et al. 1992\nocite{GelfandSmithLee92}; Karabatsos \&\ Walker, 2012a%
\nocite{KarabatsosWalker12c}).

Finally, the software implements Rao-Blackwellization (RB)\ methods (Gelfand
\&\ Mukhopadhyay, 1995\nocite{GelfandMukhopadhyay95}) to compute estimates
of the linear posterior predictive functionals from (\ref{pp all}) and (\ref%
{pd all}). In contrast, the quantile functional $Q_{n}(u\,|\,\mathbf{x})$ is
estimated from order statistics of MCMC\ samples from the posterior
predictive distribution of $Y$\ given $\mathbf{x}$. The 95\%\ posterior
credible interval of the quantile functional $Q(u\,|\,\mathbf{x})$ can be
viewed in a PP-plot (Wilk \&\ Gnanadesikan, 1968\nocite{WilkGnanadesikan68})
of the 95\%\ posterior interval of the c.d.f. $F(u\,|\,\mathbf{x})$, using
available software menu options. The hazard functional $H_{n}(y\,|\,\mathbf{x%
})$ and the cumulative hazard functional $\Lambda _{n}(y\,|\,\mathbf{x})$
are derived from RB estimates of the linear functionals $f_{n}(y\,|\,\mathbf{%
x})$ and $F_{n}(y\,|\,\mathbf{x})$. The same is true for the
partial-dependence functionals $Q_{n}(u\,|\,\mathbf{x}_{\mathcal{S}})$, $%
H_{n}(y\,|\,\mathbf{x}_{\mathcal{S}})$, and $\Lambda _{n}(y\,|\,\mathbf{x}_{%
\mathcal{S}})$.

\section{Key BNP Regression Models}

A BNP infinite-mixture\ regression model has the general form:%
\begin{equation}
f_{G_{\mathbf{x}}}(y\,|\,\mathbf{x;\,}\boldsymbol{\zeta })=\dint f(y\,|\,%
\mathbf{x};\boldsymbol{\psi },\boldsymbol{\theta }(\mathbf{x}))\mathrm{d}G_{%
\mathbf{x}}(\boldsymbol{\theta })=\dsum\limits_{j=1}^{\infty }\,f(y\,|\,%
\mathbf{x};\boldsymbol{\psi },\boldsymbol{\theta }_{j}(\mathbf{x}))\omega
_{j}(\mathbf{x}),  \label{BNP regression}
\end{equation}%
given a covariate ($\mathbf{x}$) dependent, discrete mixing distribution $G_{%
\mathbf{x}}$; kernel (component) densities\newline
$f(y\,|\,\mathbf{x};\boldsymbol{\psi },\boldsymbol{\theta }_{j}(\mathbf{x}))$
with component indices $j=1,2,\ldots $, respectively; with fixed parameters $%
\boldsymbol{\psi }$; and with component parameters $\boldsymbol{\theta }_{j}(%
\mathbf{x})$ having sample space $\Theta $; and given mixing weights $%
(\omega _{j}(\mathbf{x}))_{j=1}^{\infty }$ that sum to 1 at every $\mathbf{x}%
\in \mathcal{X}$, with $\mathcal{X}$ the covariate space.

In the infinite-mixture model (\ref{BNP regression}), the
covariate-dependent mixing distribution is a random probability measure that
has the general form\footnote{%
Throughout, $\delta _{\boldsymbol{\theta }}(\cdot )$ denotes a degenerate
probability measure (distribution) with point mass at $\boldsymbol{\theta }$%
, such that $\boldsymbol{\theta }^{\ast }\sim \delta _{\boldsymbol{\theta }}$
and $\Pr (\boldsymbol{\theta }^{\ast }=\boldsymbol{\theta })=1$. Also, $%
\delta _{\boldsymbol{\theta }}(B)=1$ if $\boldsymbol{\theta }\in B$ and $%
\delta _{\boldsymbol{\theta }}(B)=0$ if $\boldsymbol{\theta }\notin B$, for $%
\forall B\in \mathcal{B}(\Theta )$.},%
\begin{equation}
G_{\mathbf{x}}(B)=\dsum\limits_{j=1}^{\infty }\omega _{j}(\mathbf{x})\delta
_{\boldsymbol{\theta }_{j}(\mathbf{x})}(B),\text{ \ }\forall B\in \mathcal{B}%
(\Theta ),  \label{MixDist}
\end{equation}%
and is therefore an example of a species sampling model (Pitman, 1995\nocite%
{Pitman95}).

The mixture model (\ref{BNP regression}) is completed by the specification
of a prior distribution $\Pi (\boldsymbol{\zeta })$ on the space $\Omega _{%
\boldsymbol{\zeta }}=\{\boldsymbol{\zeta }\}$ of the infinite-dimensional
model parameter, given by:%
\begin{equation}
\boldsymbol{\zeta }=(\boldsymbol{\psi },(\boldsymbol{\theta }_{j}(\mathbf{x}%
),\omega _{j}(\mathbf{x}))_{j=1}^{\infty },\mathbf{x}\in \mathcal{X}).
\label{BNP reg parameters}
\end{equation}%
The BNP\ infinite-mixture regression model (\ref{BNP regression})-(\ref%
{MixDist}), completed by the specification of a prior distribution $\Pi (%
\boldsymbol{\zeta })$, is very general and encompasses, as special cases:
fixed- and random-effects linear and generalized linear models (McCullagh
\&\ Nelder, 1989\nocite{McCullaghNelder89}; Verbeke \&\ Molenberghs, 2000%
\nocite{VerbekeMolenberghs2000}; Molenberghs \&\ Verbeke, 2005\nocite%
{MolenberghsVerbeke05}), finite-mixture latent-class and hierarchical
mixtures-of-experts regression models (McLachlan \&\ Peel, 2000\nocite%
{McLachlanPeel00}; Jordan \&\ Jacobs, 1994\nocite{JordanJacobs94}), and
infinite-mixtures of Gaussian process regressions (Rasmussen \&\ Ghahramani,
2002\nocite{RasmussenGhahramani02}).

In the general BNP\ model (\ref{BNP regression})-(\ref{MixDist}), assigned
prior $\Pi (\boldsymbol{\zeta })$, the kernel densities $f(y\,|\,\mathbf{x};%
\boldsymbol{\psi },\boldsymbol{\theta }_{j}(\mathbf{x}))$ may be specified
as covariate independent,\ with: $f(y\,|\,\mathbf{x};\boldsymbol{\psi },%
\boldsymbol{\theta }_{j}(\mathbf{x})):=f(y\,|\,\boldsymbol{\psi },%
\boldsymbol{\theta }_{j})$; and may not contain fixed parameters $%
\boldsymbol{\psi }$, in which case $\boldsymbol{\psi }$ is null. Also for
the model, covariate dependence is not necessarily specified for the mixing
distribution, so that $G_{\mathbf{x}}:=G$. No covariate dependence is
specified for the mixing distribution if and only if both the component
parameters and the mixture weights are covariate independent, with $%
\boldsymbol{\theta }_{j}(\mathbf{x}):=\boldsymbol{\theta }_{j}$ and $\omega
_{j}(\mathbf{x}):=\omega _{j}$. The mixing distribution $G_{\mathbf{x}}$ is
covariate dependent if the component parameters $\boldsymbol{\theta }_{j}(%
\mathbf{x})$ or the mixture weights $\omega _{j}(\mathbf{x})$ are specified
as covariate dependent.

Under the assumption of no covariate dependence in the mixing distribution,
with $G_{\mathbf{x}}:=G$, the Dirichlet process (Ferguson, 1973\nocite%
{Ferguson73}) provides a standard and classical choice of BNP\ prior
distribution on the space of probability measures $\mathcal{G}_{\Theta }$ $%
=\{G\}_{\Theta }$ on the sample space $\Theta $. The Dirichlet process is
denoted $\mathcal{DP}(\alpha ,G_{0})$ with precision parameter $\alpha $ and
baseline distribution\ (measure) $G_{0}$. We denote $G\sim \mathcal{DP}%
(\alpha ,G_{0})$ when the random probability measure $G$ is assigned a $%
\mathcal{DP}(\alpha ,G_{0})$ prior distribution on $\mathcal{G}_{\Theta }$.
Under the $\mathcal{DP}(\alpha ,F_{0})$ prior, the (prior)\ mean and
variance of $G$ are given respectively by (Ferguson, 1973\nocite{Ferguson73}%
): 
\begin{subequations}
\begin{eqnarray}
\mathbb{E}[G(B)\,|\,\alpha ,G_{0}] &=&\dfrac{\alpha G_{0}(B)}{\alpha }%
=G_{0}(B),  \label{EG DP} \\
\mathbb{V}[G(B)\,|\,\alpha ,G_{0}] &=&\frac{G_{0}(B)[1-G_{0}(B)]}{\alpha +1},%
\text{ \ \ }\forall B\in \mathcal{B}(\Theta ).  \label{VarG DP}
\end{eqnarray}%
For the $\mathcal{DP}(\alpha ,G_{0})$ prior, (\ref{EG DP}) shows that the
baseline distribution $G_{0}$ represents the prior mean (expectation)\ of $%
G, $ and the prior variance of $G$ is inversely proportional to the
precision parameter $\alpha $, as shown in (\ref{VarG DP}). The variance of $%
G$ is increased (decreased, resp.) as $\alpha $ becomes smaller (larger,
resp.). In practice, a standard choice of baseline distribution $G_{0}(\cdot
)$ is provided by the normal $\mathrm{N}(\mu ,\sigma ^{2})$ distribution.
The $\mathcal{DP}(\alpha ,G_{0})$ can also be characterized in terms of a
Dirichlet ($\mathrm{Di}$) distribution. That is, if $G\sim \mathcal{DP}%
(\alpha ,G_{0})$, then: 
\end{subequations}
\begin{equation}
(G(B_{1}),\ldots ,G(B_{k}))\,|\,\alpha ,G_{0}\sim \mathrm{Di}(\alpha
G_{0}(B_{1}),\ldots ,\alpha G_{0}(B_{k})),
\end{equation}%
for every choice of $k\geq 1$ (exhaustive) partitions $B_{1},\ldots ,B_{k}$
of the sample space, $\Theta $.

The $\mathcal{DP}(\alpha ,G_{0})$ can also be characterized as a particular
"stick-breaking"\ stochastic process (Sethuraman, 1994\nocite{Sethuraman94};
Sethuraman \&\ Tiwari, 1982\nocite{SethuramanTiwari82}). A random
probability measure ($G$) that is drawn from the $\mathcal{DP}(\alpha
,G_{0}) $ prior, with $G\sim \mathcal{DP}(\alpha ,G_{0})$, is constructed by
first taking independently and identically distributed (i.i.d.)\ samples of $%
(\upsilon ,\boldsymbol{\theta })$ from the following distributions: 
\begin{subequations}
\label{SB const}
\begin{eqnarray}
&&\upsilon _{j}\,|\,\alpha \sim \mathrm{Be}(1,\alpha ),\text{ \ \ }%
j=1,2,\ldots ,  \label{betaSB} \\
&&\boldsymbol{\theta }_{j}\,|\,G_{0}\sim G_{0},\text{ \ \ }j=1,2,\ldots ,
\label{atomSB}
\end{eqnarray}%
and then using the samples $(\upsilon _{j},\boldsymbol{\theta }%
_{j})_{j=1}^{\infty }$ to construct the random probability measure by: 
\begin{equation}
G(B)=\dsum\limits_{j=1}^{\infty }\omega _{j}\delta _{\boldsymbol{\theta }%
_{j}}(B),\text{ \ \ }\forall B\in \mathcal{B}(\Theta ).  \label{G SB}
\end{equation}%
Above, the $\omega _{j}$s are mixture weights, particularly, stick-breaking
weights constructed by:%
\begin{equation}
\omega _{j}=\upsilon _{j}\prod_{l=1}^{j-1}(1-\upsilon _{l}),\text{ \ for \ }%
j=1,2,\ldots ,  \label{SB weights}
\end{equation}%
and they sum to 1 (i.e., $\tsum\nolimits_{j=1}^{\infty }\omega _{j}=1$).

More in words, a random probability measure, $G$, drawn from a $\mathcal{DP}%
(\alpha ,G_{0})$ prior distribution on $\mathcal{G}_{\Theta }$ $%
=\{G\}_{\Theta }$, can be represented as infinite-mixtures of degenerate
probability measures (distributions).\ Such a random distribution is
discrete with probability 1, which is obvious because the degenerate
probability measure ($\delta _{\boldsymbol{\theta }_{j}}(\cdot )$) is
discrete. The locations $\boldsymbol{\theta }_{j}$ of the point masses are a
sample from $G_{0}$. The random weights $\omega _{j}$ are obtained from a
stick-breaking procedure, described as follows. First, imagine a stick of
length 1. As shown in (\ref{SB weights}), at stage $j=1$ a piece is broken
from this stick, and then the value of the first weight $\omega _{1}$ is set
equal to the length of that piece, with $\omega _{1}=\upsilon _{1}$. Then at
stage $j=2$, a piece is broken from a stick of length $1-\omega _{1}$, and
then the value of the second weight $\omega _{2}=\upsilon _{2}(1-\upsilon
_{1})$ is set equal to the length of that piece. This procedure is repeated
for\ $j=1,2,3,4,\ldots $, where at any given stage $j$, a piece is broken
from a stick of length $1-\tsum\nolimits_{l=1}^{j-1}\omega _{j}$, and then
the value of the weight $\omega _{j}$ is set equal to the length of that
piece, with $\omega _{j}=\upsilon _{j}\prod_{l=1}^{j-1}(1-\upsilon _{l})$.
The entire procedure results in weights $(\omega _{j})_{j=1}^{\infty }$ that
sum to 1 (almost surely).

The stick-breaking construction (\ref{SB const}) immediately suggests
generalizations of the $\mathcal{DP}(\alpha ,G_{0})$, especially by means of
increasing the flexibility of the prior (\ref{betaSB}) for the random
parameters $(\upsilon _{j})_{j=1}^{\infty }$ that construct the
stick-breaking mixture weights (\ref{SB weights}). One broad generalization
is given by a general stick-breaking process (Ishwaran \&\ James, 2001\nocite%
{IshwaranJames01}), denoted $\mathcal{SB}(\mathbf{a},\mathbf{b},G_{0})$ with
positive parameters $\mathbf{a}=(a_{1},a_{2},\ldots )$ and $\mathbf{b}%
=(b_{1},b_{2},\ldots )$, which gives a prior on $\mathcal{G}_{\Theta }$ $%
=\{G\}_{\Theta }$. This process replaces the i.i.d. beta distribution
assumption in (\ref{betaSB}), with the more general assumption of
independent beta distributions, with 
\end{subequations}
\begin{equation}
\upsilon _{j}\,|\,\alpha \sim \mathrm{Be}(a_{j},b_{j}),\text{ \ for }%
j=1,2,\ldots \text{.}  \label{SPpriorbeta}
\end{equation}%
In turn, there are many interesting special cases of the $\mathcal{SB}(%
\mathbf{a},\mathbf{b},G_{0})$ process prior, including:

\begin{enumerate}
\item The Pitman-Yor (Poisson-Dirichlet) process, denoted $\mathcal{PY}%
(a,b,G_{0})$, which assumes $a_{j}=1-a$ and $b_{j}=b+ja$, for $j=1,2,\ldots $
, in (\ref{SPpriorbeta}), with $0\leq a<1$ and $b>-a$ (Perman, et al. 1992%
\nocite{Perman_Pitman_Yor_92}; Pitman \&\ Yor, 1997\nocite{PitmanYor97}).

\item The beta two-parameter process, which assumes $a_{j}=a$ and $b_{j}=b$
in (\ref{SPpriorbeta}) (Ishwaran \&\ Zarepour, 2000\nocite%
{IshwaranZarepour00}).

\item The normalized stable process (Kingman, 1975\nocite{Kingman75}), which
is equivalent to the $\mathcal{PY}(a,0,G_{0})$ process, with $0\leq a<1$ and 
$b=0$.

\item The Dirichlet process $\mathcal{DP}(\alpha ,G_{0})$, which assumes $%
a_{j}=1$ and $b_{j}=\alpha $ in (\ref{SPpriorbeta}), and with is equivalent
to the $\mathcal{PY}(0,\alpha ,G_{0})$ process.

\item The geometric weights prior, denoted $\mathcal{GW}(a,b,G_{0})$, which
assumes in (\ref{SPpriorbeta}) the equality restriction $\upsilon =\upsilon
_{j}$ for $j=1,2,\ldots ,$ leading to mixture weights (\ref{SB weights})
that can be re-written as $\omega _{j}=\upsilon \left( 1-\upsilon \right)
^{j-1}$, for $j=1,2,\ldots $ (Fuentes-Garc\'{\i}a, et al. 2009, 2010\nocite%
{Fuentes-GarciaMenaWalker09}\nocite{Fuentes-GarciaMenaWalker10}). These
mixture weights may be assigned a beta prior distribution, with $\upsilon
\sim \mathrm{Be}(a,b)$.
\end{enumerate}

\noindent Another generalization of the $\mathcal{DP}(\alpha ,G_{0})$ is
given by the mixture of Dirichlet process (MDP), defined by the
stick-breaking construction (\ref{SB const}), after sampling from prior
distributions $\alpha \sim \Pi \left( \alpha \right) $ and $\boldsymbol{%
\vartheta }\sim \Pi \left( \boldsymbol{\vartheta }\right) $ for the
precision and baseline parameters (Antoniak, 1974\nocite{Antoniak74}).

A BNP\ prior distribution on $\mathcal{G}_{\Theta }$ $=\{G\}_{\Theta }$,
defined by a Normalized Random Measure (NRM) process, assumes that a
discrete random probability measure $G$, given by (\ref{G SB}), is
constructed by mixture weights that have the form: 
\begin{equation}
\omega _{j}=\dfrac{I_{j}}{\tsum\nolimits_{l=1}^{\infty }I_{l}},\text{ \ }%
j=1,2,\ldots ;\text{ \ }\omega _{j}\geq 0,\text{ }\tsum\nolimits_{l=1}^{%
\infty }\omega _{l}=1.
\end{equation}%
The $I_{1},I_{2},I_{3},\ldots $ are the jump sizes of a non-Gaussian L\'{e}%
vy process whose sum is almost surely finite (see e.g. James et al., 2009%
\nocite{JamesLijoiPrunster09}), and are therefore stationary independent
increments (Bertoin, 1998\nocite{Bertoin98}). The $\mathcal{DP}(\alpha
,G_{0})$\ is a special NRM\ process which assumes that $\tsum%
\nolimits_{j=1}^{\infty }I_{j}\sim \mathrm{Ga}(\alpha ,1)$ (Ferguson, 1973%
\nocite{Ferguson73}, pp. 218-219).

An important NRM is given by the normalized inverse-Gaussian $\mathcal{NIG}%
(c,G_{0})$ process (Lijoi et al., 2005\nocite{LijoiMenaPrunster05}), which
can be characterized as a stick-breaking process (Favaro, et al., 2012\nocite%
{FavaroLijoiPrunster12}), defined by the stick-breaking construction (\ref%
{SB const}), after relaxing the i.i.d. assumption (\ref{betaSB}), by
allowing for dependence among the $\upsilon _{j}$ distributions, with: 
\begin{subequations}
\label{NIG process stick}
\begin{eqnarray}
\upsilon _{j} &=&\dfrac{\upsilon _{1j}}{\upsilon _{1j}+\upsilon _{0j}},\text{
\ }j=1,2,\ldots  \label{favaro pdf} \\
\upsilon _{1j} &\sim &\mathrm{GIG}(c^{2}/\{\tprod%
\nolimits_{l=1}^{j-1}(1-V_{l})\}^{\mathbf{1}(j>1)},1,-j/2), \\
\upsilon _{0j} &\sim &\mathrm{IG}(1/2,2).
\end{eqnarray}%
The random variables (\ref{favaro pdf}) follow normalized generalized
inverse-Gaussian distributions, with p.d.f. given by equation (4)\ in
Favaro, et al. (2012\nocite{FavaroLijoiPrunster12}).

Stick-breaking process priors can be characterized in terms of the
clustering behavior that it induces in the posterior predictive distribution
of $\boldsymbol{\theta }$. Let $\{\boldsymbol{\theta }_{c}^{\ast
}:c=1,\ldots ,k_{n}\leq n\}$ be the $k_{n}\leq n$ unique values (clusters)
among the $n$ observations of a data set. Let $\Upsilon _{n}=\{\mathcal{C}%
_{1},\ldots ,\mathcal{C}_{c},\ldots ,\mathcal{C}_{k_{n}}\}$ be the random
partition of the integers $\{1,\ldots ,n\}$. Each cluster is defined by $%
\mathcal{C}_{c}=\{i:\boldsymbol{\theta }_{i}=\boldsymbol{\theta }_{c}^{\ast
}\}\subset \{1,\ldots ,n\},$ and has size $n_{c}=|\mathcal{C}_{c}|$, with
cluster frequency counts $\mathbf{n}_{n}=(n_{1},\ldots ,n_{c},\ldots
,n_{k_{n}})$ and $\tsum\nolimits_{c=1}^{k_{n}}n_{c}=n$.

When $G$\ is assigned a Pitman-Yor $\mathcal{PY}(a,b,G_{0})$ process prior,
the posterior predictive probability of a new observation $\boldsymbol{%
\theta }_{n+1}$ is defined by: 
\end{subequations}
\begin{equation}
P(\boldsymbol{\theta }_{n+1}\in B\,|\,\,\boldsymbol{\theta }_{1},\ldots ,%
\boldsymbol{\theta }_{n})=\dfrac{b+ak_{n}}{b+n}G_{0}(B)+\dsum%
\limits_{c=1}^{k_{n}}\dfrac{n_{c}-a}{b+n}\delta _{\boldsymbol{\theta }%
_{c}^{\ast }}(B),\text{ \ \ }\forall B\in \mathcal{B}(\Theta ).
\label{PY pred}
\end{equation}%
That is, $\boldsymbol{\theta }_{n+1}$ forms a new cluster with probability $%
(b+ak_{n})/(b+n)$, and otherwise with probability $(n_{c}-a)/(b+n)$, $%
\boldsymbol{\theta }_{n+1}$ is allocated to old cluster $\mathcal{C}_{c}$,
for $c=1,\ldots ,k_{n}$. Recall that the normalized stable process (Kingman,
1975\nocite{Kingman75}) is equivalent to the $\mathcal{PY}(a,0,G_{0})$
process with $0\leq a<1$ and $b=0$; and the $\mathcal{DP}(\alpha ,G_{0})$ is
the $\mathcal{PY}(0,b,G_{0})$ process with $a=0$ and $b=\alpha $.

Under the $\mathcal{NIG}(c,G_{0})$ process prior, the posterior predictive
distribution is defined by the probability function, 
\begin{subequations}
\label{NIG all}
\begin{equation}
P(y_{n+1}\in B\,|\,\,\boldsymbol{\theta }_{1},\ldots ,\boldsymbol{\theta }%
_{n})=w_{0}^{(n)}G_{0}(B)+w_{1}^{(n)}\dsum\limits_{c=1}^{k_{n}}(n_{c}-.5)%
\delta _{y_{c}^{\ast }}(B),\text{ \ \ }\forall B\in \mathcal{B}(\Theta ),
\label{GG pred}
\end{equation}%
with:\noindent 
\begin{equation}
w_{0}^{(n)}=\dfrac{\tsum\limits_{l=0}^{n}\dbinom{n}{l}(-c^{2})^{-l+1}\Gamma
(k_{n}+1+2l-2n\,;\,c)}{2n\tsum\limits_{l=0}^{n-1}\dbinom{n-1}{l}%
(-c^{2})^{-l}\Gamma (k_{n}+2+2l-2n\,;\,c)},  \label{w0 GG}
\end{equation}%
\begin{equation}
w_{1}^{(n)}=\dfrac{\tsum\limits_{l=0}^{n}\dbinom{n}{l}(-c^{2})^{-l+1}\Gamma
(k_{n}+2l-2n\,;\,c)}{n\tsum\limits_{l=0}^{n-1}\dbinom{n-1}{l}%
(-c^{2})^{-l}\Gamma (k_{n}+2+2l-2n\,;\,c)},  \label{w1 GG}
\end{equation}%
where $\Gamma (\cdot \,;\,\cdot )$ is the incomplete gamma function (Lijoi
et al., 2005, p. 1283\nocite{LijoiMenaPrunster05}). Finally, exchangeable
partition models (e.g., Hartigan, 1990\nocite{Hartigan90}; Barry \&\
Hartigan, 1993\nocite{BarryHartigan93}; Quintana \&\ Iglesias, 2003\nocite%
{QuintanaIglesias03}) also give rise to random clustering structures of a
form (\ref{NIG all}), and therefore coincide with the family of Gibbs-type
priors, which include the $\mathcal{PY}(a,b,G_{0})$ and $\mathcal{NIG}%
(c,G_{0})$ processes and their special cases. More detailed discussions on
the clustering behavior induced by various BNP\ priors are given by De\
Blasi, et al. (2015\nocite{DeBlasiEtAl15}).

So far, we have described only BNP\ priors for the mixture distribution (\ref%
{MixDist}) of the general BNP\ regression model (\ref{BNP regression}),
while assuming no covariate dependence in the mixing distribution, with $G_{%
\mathbf{x}}:=G$. We now consider dependent BNP\ processes. A seminal example
is given by the Dependent Dirichlet process ($\mathcal{DDP}(\alpha _{\mathbf{%
x}},G_{0\mathbf{x}})$) (MacEachern 1999\nocite{MacEachern99}; 2000\nocite%
{MacEachern00}; 2001\nocite{MacEachern01}), which models a covariate ($%
\mathbf{x}$) dependent process $G_{\mathbf{x}}$, by allowing either the
baseline distribution $G_{0\mathbf{x}}$, the stick-breaking mixture weights $%
\omega _{j}(\mathbf{x})$, and/or the precision parameter $\alpha _{\mathbf{x}%
}$ to depend on covariates $\mathbf{x}$. In general terms, a random
dependent probability measure $G_{\mathbf{x}}\,|\,\alpha _{\mathbf{x}},G_{0%
\mathbf{x}}\sim \mathcal{DDP}(\alpha _{\mathbf{x}},G_{0\mathbf{x}})$ can be
represented by Sethuraman's (1994\nocite{Sethuraman94})\ stick-breaking
construction, as:%
\begin{eqnarray}
G_{\mathbf{x}}(B) &=&\tsum\nolimits_{j=1}^{\infty }\omega _{j}(\mathbf{x}%
)\delta _{\theta _{j}(\mathbf{x})}(B),\text{ \ \ }\forall B\in \mathcal{B}%
(\Theta ), \\
\omega _{j}(\mathbf{x}) &=&\upsilon _{j}(\mathbf{x})\tprod%
\nolimits_{k=1}^{j-1}(1-\upsilon _{k}(\mathbf{x})), \\
\upsilon _{j}(\mathbf{x}) &:&\mathcal{X}\rightarrow \lbrack 0,1], \\
\upsilon _{j} &\sim &Q_{\mathbf{x}j}, \\
\theta _{j}(\mathbf{x}) &\sim &G_{0\mathbf{x}}.
\end{eqnarray}%
\noindent Next, we describe an important BNP\ regression model, with a
dependent mixture distribution $G_{\mathbf{x}}$ assigned a specific $%
\mathcal{DDP}(\alpha _{\mathbf{x}},G_{0\mathbf{x}})$ prior.

\subsection{ANOVA-Linear DDP\ model}

Assume that the data $\mathcal{D}_{n}=\{(y_{i},\mathbf{x}_{i})\}_{i=1}^{n}$
can be stratified into $N_{h}$ groups, indexed by $h=1,\ldots ,N_{h}$,
respectively. For each of group $h$, let $y_{i(h)}$ be the $i$th dependent
observation of group $h$, and let $\mathbf{y}%
_{h}=(y_{i(h)})_{i(h)=1}^{n_{h}} $ be the column vector of $n_{h}$ dependent
observations, corresponding to an observed design matrix $\mathbf{X}_{h}=(%
\mathbf{x}_{1(h)}^{\mathbf{\intercal }},\ldots ,\mathbf{x}_{i(h)}^{\mathbf{%
\intercal }},\ldots ,\mathbf{x}_{n_{h}}^{\mathbf{\intercal }})$ of $n_{h}$
rows of covariate vectors $\mathbf{x}_{i(h)}^{\mathbf{\intercal }}$
respectively. Possibly, each of the $N_{h}$ groups of observations has only
one observation (i.e., $n_{h}=1$), in which case $N_{h}=n$.

The ANOVA-linear DDP model (De Iorio, et al. 2004\nocite%
{DeIorioMullerRosnerMacEachern04}; M\"{u}ller et al., 2005\nocite%
{MullerRosnerDeIorioMacEachern05}) can be defined as: 
\end{subequations}
\begin{subequations}
\label{ANOVA DDP}
\begin{eqnarray}
(y_{i(h)})_{i(h)=1}^{n_{h}}\,|\,\mathbf{X}_{h} &\sim &f(\mathbf{y}_{h}\,|\,%
\mathbf{X}_{h};\boldsymbol{\zeta }),\text{ \ }h=1,\ldots ,N_{h} \\
f(\mathbf{y}_{h}\,|\,\mathbf{X}_{h};\boldsymbol{\zeta })
&=&\dsum\limits_{j=1}^{\infty }\left\{ \tprod\limits_{i(h)=1}^{n_{h}}\text{%
\textrm{n}}(y_{i(h)}\,|\,\mathbf{x}_{i(h)}^{\mathbf{\intercal }}\boldsymbol{%
\beta }_{j},\sigma ^{2})\right\} \omega _{j} \\
\omega _{j} &=&\upsilon _{j}\tprod\nolimits_{l=1}^{j-1}(1-\upsilon _{l})
\label{SB weights DDP} \\
\upsilon _{j}\,|\,\alpha &\sim &\text{\textrm{Be}}(1,\alpha ) \\
\boldsymbol{\beta }_{j}\,|\,\boldsymbol{\mu },\boldsymbol{T} &\sim &\text{$%
\mathrm{N}$}(\boldsymbol{\mu },\boldsymbol{T}) \\
\sigma ^{2} &\sim &\text{\textrm{IG}}(a_{0}/2,a_{0}/2) \\
\boldsymbol{\mu },\boldsymbol{T} &\sim &\text{$\mathrm{N}$}(\boldsymbol{\mu }%
\,|\,\mathbf{0},r_{0}\mathbf{I}_{p+1})\text{\textrm{IW}}(\boldsymbol{T}%
\,|\,p+3,s_{0}\mathbf{I}_{p+1}) \\
\alpha &\sim &\text{$\mathrm{Ga}$}(a_{\alpha },b_{\alpha }).
\end{eqnarray}%
Therefore, all the model parameters are assigned prior distributions, which
together, define the joint prior p.d.f. for $\boldsymbol{\zeta }\in \Omega _{%
\boldsymbol{\zeta }}$ by: 
\end{subequations}
\begin{subequations}
\label{ANOVA DDP prior pdf}
\begin{eqnarray}
\pi (\boldsymbol{\zeta }) &=&\dprod\limits_{j=1}^{\infty }\text{\textrm{be}}%
(\upsilon _{j}\,|\,1,\alpha )\mathrm{n}(\boldsymbol{\beta }_{j}\,|\,%
\boldsymbol{\mu },\boldsymbol{T})\text{\textrm{ig}}(\sigma
^{2}\,|\,a_{0}/2,a_{0}/2) \\
&&\times \mathrm{n}(\boldsymbol{\mu }\,|\,\mathbf{0},r_{0}\mathbf{I}_{p+1})%
\text{\textrm{iw}}(\boldsymbol{T}\,|\,p+3,s_{0}\mathbf{I}_{p+1})\mathrm{g}%
\text{$\mathrm{a}$}(\alpha \,|\,a_{\alpha },b_{\alpha }).
\end{eqnarray}%
\noindent As shown, the ANOVA-linear DDP\ model (\ref{ANOVA DDP}) is based
on a mixing distribution $G(\boldsymbol{\beta })$ assigned a $\mathcal{DP}%
(\alpha ,G_{0})$ prior, with precision parameter $\alpha $ and multivariate
normal baseline distribution, $G_{0}(\cdot ):=\mathrm{N}$$(\cdot \,|\,%
\boldsymbol{\mu },\boldsymbol{T})$. Prior distributions are assigned to $%
(\alpha ,\boldsymbol{\mu },\boldsymbol{T})$ in order to allow for posterior
inferences to be robust to different choices of the $\mathcal{DP}(\alpha
,G_{0})$ prior parameters.

The ANOVA-linear DDP\ model (\ref{ANOVA DDP}) is equivalent to the BNP\
regression model (\ref{BNP regression}), with normal kernel densities 
\textrm{n}$(y_{i}\,|\,\mu _{j},\sigma ^{2})$ and mixing distribution $G_{%
\mathbf{x}}(\mu )$ (\ref{MixDist}) assigned a $\mathcal{DDP}(\alpha ,G_{0%
\mathbf{x}})$ prior, where: 
\end{subequations}
\begin{equation}
G_{\mathbf{x}}(B)=\tsum\nolimits_{j=1}^{\infty }\omega _{j}\delta _{\mathbf{x%
}^{\mathbf{\intercal }}\boldsymbol{\beta }\newline
}(B),\ \forall B\in \mathcal{B}(\Theta ),  \label{depmix}
\end{equation}%
with $\boldsymbol{\beta }_{j}\,|\,\boldsymbol{\mu },\boldsymbol{T}\sim 
\mathrm{N}$$(\boldsymbol{\mu },\boldsymbol{T})$ and $\sigma ^{2}\sim $ 
\textrm{IG}$(a_{0}/2,a_{0}/2)$ (i.e., $G_{0}(\cdot )=\mathrm{N}$$(%
\boldsymbol{\beta }\,|\boldsymbol{\mu },\boldsymbol{T})$\textrm{IG}$(\sigma
^{2}\,|a_{0}/2,a_{0}/2)$), and with the $\omega _{j}$ stick-breaking weights
(\ref{SB weights DDP}) (De Iorio, et al. 2004\nocite%
{DeIorioMullerRosnerMacEachern04}).

A menu option in the \textbf{Bayesian regression} software labels the
ANOVA-linear DDP\ model (\ref{ANOVA DDP}) as the "Dirichlet process mixture
of homoscedastic linear regressions model" (for Step 8 of a data analysis;
see next section). The software allows the user to analyze data using any
one of many variations of the model (\ref{ANOVA DDP}). Variations of this
DDP\ model include: "mixture of linear regressions" models, as labeled by a
menu option of the software, with mixing distribution $G(\boldsymbol{\beta }%
,\sigma ^{2})$ for the coefficients and the error variance parameters;
"mixture of random intercepts" models, with mixture distribution $G(\beta
_{0})$ for only the intercept parameter $\beta _{0},$ and with independent
normal priors for the slope coefficient parameters $(\beta _{k})_{k=1}^{p}$;
mixture models having mixture distribution $G$ assigned either a Pitman-Yor $%
\mathcal{PY}(a,b,G_{0})$ (including the normalized stable process prior),
beta process, geometric weights, or normalized inverse-Gaussian process $%
\mathcal{NIG}(c,G_{0})$ prior, each implying, respectively, a dependent BNP\
prior for a covariate-dependent mixing distribution (\ref{depmix}) (using
similar arguments made for the DDP\ model by De Iorio, et al. 2004\nocite%
{DeIorioMullerRosnerMacEachern04}); and mixed-logit or mixed-probit
regression models for a binary (0 or 1) or ordinal ($c=0,1,\ldots ,m$)
dependent variable. Also, suppose that the ANOVA-linear DDP\ model (\ref%
{ANOVA DDP}) is applied to time-lagged dependent variable data (which can be
set up using a menu option in the software; see Section 4.1, Step 3, and
Section 4.3). Then this model is defined by an infinite-mixture of
autoregressions, with mixture distribution assigned a time-dependent DDP (Di
Lucca, et al. 2012\nocite{DiLucca_etal12}). The Help menu of the software
provides a full list of models that are available from the software.

\subsection{Infinite-Probits Mixture Linear Model}

As mentioned, typical BNP infinite-mixture models assume that the mixture
weights have the stick-breaking form (\ref{SB weights}). However, a BNP\
model may have weights with a different form. The infinite-probits model is
a Bayesian nonparametric regression model (\ref{BNP regression})-(\ref%
{MixDist}), with prior $\Pi (\boldsymbol{\zeta })$, and with mixture
distribution (\ref{MixDist}) defined by a dependent normalized random
measure (Karabatsos \& Walker, 2012\nocite{KarabatsosWalker12c}).

For data, $\mathcal{D}_{n}=\{(y_{i},\mathbf{x}_{i})\}_{i=1}^{n}$, a Bayesian
infinite-probits mixture model can be defined by: 
\begin{subequations}
\label{IP mixture reg model}
\begin{eqnarray}
y_{i}\,|\,\mathbf{x}_{i} &\sim &f(y\,|\,\mathbf{x}_{i};\boldsymbol{\zeta }),%
\text{ \ }i=1,\ldots ,n \\
f(y\,|\,\mathbf{x};\boldsymbol{\zeta }) &=&\tsum\limits_{j=-\infty }^{\infty
}\text{\textrm{n}}(y\,|\,\mu _{j}+\mathbf{x}^{\mathbf{\intercal }}%
\boldsymbol{\beta },\sigma ^{2})\omega _{j}(\mathbf{x})
\label{IP inf mixture} \\
\omega _{j}(\mathbf{x}) &=&\Phi \left( \dfrac{j-\mathbf{x}^{\mathbf{%
\intercal }}\boldsymbol{\beta }_{\omega }}{\sigma _{\omega }}\right) -\Phi
\left( \dfrac{j-1-\mathbf{x}^{\mathbf{\intercal }}\boldsymbol{\beta }%
_{\omega }}{\sigma _{\omega }}\right)  \label{IP mix weights} \\
\mu _{j}\,|\,\sigma _{\mu }^{2} &\sim &\text{$\mathrm{N}$}(0,\sigma _{\mu
}^{2}) \\
\sigma _{\mu } &\sim &\text{\textrm{U}}(0,b_{\sigma \mu }) \\
\beta _{0}\,|\,\sigma ^{2} &\sim &\text{$\mathrm{N}$}(0,\sigma ^{2}v_{\beta
0}\rightarrow \infty ) \\
\beta _{k}\,|\,\sigma ^{2} &\sim &\text{$\mathrm{N}$}(0,\sigma ^{2}v),\text{ 
}k=1,\ldots ,p \\
\sigma ^{2} &\sim &\text{\textrm{IG}}(a_{0}/2,a_{0}/2) \\
\boldsymbol{\beta }_{\omega }\,|\,\sigma _{\omega }^{2} &\sim &\text{$%
\mathrm{N}$}(\mathbf{0},\sigma _{\omega }^{2}v_{\omega }\mathbf{I}) \\
\sigma _{\omega }^{2} &\sim &\text{\textrm{IG}}(a_{\omega }/2,a_{\omega }/2),
\end{eqnarray}%
with $\Phi \left( \cdot \right) $ the normal \textrm{N}$(0,1)$ c.d.f., and
model parameters $\boldsymbol{\zeta }=((\mu _{j})_{j=1}^{\infty },\sigma
_{\mu }^{2},\boldsymbol{\beta },\sigma ^{2},\boldsymbol{\beta }_{\omega
},\sigma _{\omega })$ assigned a prior $\Pi (\boldsymbol{\zeta })$ with
p.d.f.: 
\end{subequations}
\begin{subequations}
\label{IP prior pdf}
\begin{eqnarray}
\pi (\boldsymbol{\zeta }) &=&\dprod\limits_{j=-\infty }^{\infty }\mathrm{n}%
(\mu _{j}\,|\,0,\sigma _{\mu }^{2})\text{\textrm{u}}(\sigma _{\mu
}\,|\,0,b_{\sigma \mu })\mathrm{n}(\boldsymbol{\beta }\,|\,\mathbf{0},\sigma
^{2}\mathrm{diag}(v_{\beta 0}\rightarrow \infty ,v\mathbf{J}_{p})) \\
&&\times \text{\textrm{ig}}(\sigma ^{2}\,|\,a_{0}/2,a_{0}/2)\mathrm{n}(%
\boldsymbol{\beta }_{\omega }\,|\,\mathbf{0},\sigma _{\omega }^{2}v_{\omega }%
\mathbf{I}_{p+1})\mathrm{ig}(\sigma _{\omega }^{2}\,|\,a_{\omega
}/2,a_{\omega }/2),
\end{eqnarray}%
where $\mathbf{J}_{p}$ denotes a $p\times 1$ vector of $1$s.

The \textbf{Bayesian Regression} software labels the BNP\ model (\ref{IP
mixture reg model}) as the "Infinite homoscedastic probits regression
model," in a menu option (in Step 8 of a data analysis; see next section).
This model is defined by a highly-flexible robust linear model, an infinite
mixture of linear regressions (\ref{IP inf mixture}), with random intercept
parameters $\mu _{j}$ modeled by infinite covariate-dependent mixture
weights (\ref{IP mix weights}). The model (\ref{IP mixture reg model}) has
been extended and applied to prediction analysis (Karabatsos \&\ Walker,
2012a\nocite{KarabatsosWalker12c}), meta-analysis (Karabatsos et al. 2015%
\nocite{KarabatsosTalbottWalker15}), (test)\ item-response analysis
(Karabatsos, 2015\nocite{Karabatsos15}), and causal analysis (Karabatsos \&\
Walker, 2015\nocite{KarabatsosWalker15}).

The covariate-dependent mixture weights $\omega _{j}(\mathbf{x})$ in (\ref%
{IP mix weights}), defining the mixture distribution (\ref{MixDist}), are
modeled by a probits regression for ordered categories $j=\ldots ,-2,-1,0,$ $%
1,2,\ldots $, with latent location parameter $\mathbf{x}^{\mathbf{\intercal }%
}\boldsymbol{\beta }_{\omega }$, and with latent standard deviation $\sigma
_{\omega }$ that controls the level of modality of the conditional p.d.f. $%
f(y\,|\,\mathbf{x};\boldsymbol{\zeta })$ of the dependent variable $Y$.
Specifically, as $\sigma _{\omega }\rightarrow 0$, the conditional p.d.f. $%
f(y\,|\,\mathbf{x};\boldsymbol{\zeta })$ becomes more unimodal. As $\sigma
_{\omega }$ gets larger, $f(y\,|\,\mathbf{x};\boldsymbol{\zeta })$ becomes
more multimodal (see Karabatsos \&\ Walker, 2012a\nocite{KarabatsosWalker12c}%
).

The \textbf{Bayesian Regression} software allows the user to analyze data
using any one of several versions of the infinite-probits\ regression model (%
\ref{IP mixture reg model}). Versions include models where the kernel
densities are instead specified by covariate independent normal densities 
\textrm{n}$(y\,|\,\mu _{j},\sigma _{j}^{2})$, and the mixture weights are
modeled by: 
\end{subequations}
\begin{equation}
\omega _{j}(\mathbf{x})=\Phi \left( \dfrac{j-\mathbf{x}^{\mathbf{\intercal }}%
\boldsymbol{\beta }_{\omega }}{\sqrt{\exp (\mathbf{x}^{\mathbf{\intercal }}%
\boldsymbol{\lambda }_{\omega })}}\right) -\Phi \left( \dfrac{j-\mathbf{x}^{%
\mathbf{\intercal }}\boldsymbol{\beta }_{\omega }-1}{\sqrt{\exp (\mathbf{x}^{%
\mathbf{\intercal }}\boldsymbol{\lambda }_{\omega })}}\right) ,\text{ \ \
for }j=0,\pm 1,\pm 2,\ldots ;  \label{new IP mix weights}
\end{equation}%
include models where either the individual regression coefficients $%
\boldsymbol{\beta }$ in the kernels, or the individual regression
coefficients $(\boldsymbol{\beta }_{\omega },\boldsymbol{\lambda }_{\omega })
$ in the mixture weights (\ref{new IP mix weights}) are assigned
spike-and-slab priors using the SSVS method (George \& McCulloch, 1993, 1997%
\nocite{GeorgeMcCulloch93}\nocite{GeorgeMcCulloch97}), to enable automatic
variable (covariate)\ selection inferences from the posterior distribution;
and include mixed-probit regression models for binary (0 or 1) or ordinal ($%
c=0,1,\ldots ,m$) dependent variables, each with inverse-link function
c.d.f. modeled by a covariate-dependent, infinite-mixture of normal
densities (given by (\ref{IP inf mixture}), but instead for the continuous
underlying latent dependent variable; see Karabatsos \&\ Walker, 2015\nocite%
{KarabatsosWalker15}).

\subsection{Some Linear Models}

We briefly review two basic Bayesian normal linear models from standard
textbooks (e.g., O'Hagan \&\ Forster, 2004\nocite{OhaganForster04}; Denison
et al. 2003\nocite{DenisonHolmesMallickSmith02}).

First, the Bayesian normal linear model, assigned a (conjugate)\ normal
inverse-gamma prior distribution to the coefficients and error variance
parameters, $(\boldsymbol{\beta },\sigma ^{2})$, is defined by: 
\begin{subequations}
\label{Bayesian lm}
\begin{eqnarray}
y_{i}\,|\,\mathbf{x}_{i} &\sim &f(y\,|\,\mathbf{x}_{i}),\text{ \ \ \ }%
i=1,\ldots ,n \\
f(y\,|\,\mathbf{x}) &=&\text{\textrm{n}}(y\,|\,\mathbf{x}^{\intercal }%
\boldsymbol{\beta },\sigma ^{2}) \\
\beta _{0}\,|\,\sigma ^{2} &\sim &\text{$\mathrm{N}$}(0,\sigma ^{2}v_{\beta
0}\rightarrow \infty ) \\
\beta _{k}\,|\,\sigma ^{2} &\sim &\text{$\mathrm{N}$}(0,\sigma ^{2}v_{\beta
}),\text{ \ }k=1,\ldots ,p \\
\sigma ^{2} &\sim &\text{\textrm{IG}}(a_{0}/2,a_{0}/2).
\end{eqnarray}%
An extension of the model (\ref{Bayesian lm}) is provided by the Bayesian
2-level normal random-effects model (HLM). Again, let the data $\mathcal{D}%
_{n}=\{(y_{i},\mathbf{x}_{i})\}_{i=1}^{n}$ be stratified into $N_{h}$
groups, indexed by $h=1,\ldots ,N_{h}$. Also, for each group $h$, let $%
y_{i(h)}$ be the $i$th dependent observation, and let $\mathbf{y}%
_{h}=(y_{i(h)})_{i(h)=1}^{n_{h}}$ be the column vector of $n_{h}$ dependent
observations, corresponding to an observed design matrix $\mathbf{X}_{h}=(%
\mathbf{x}_{1(h)}^{\mathbf{\intercal }},\ldots ,\mathbf{x}_{i(h)}^{\mathbf{%
\intercal }},\ldots ,\mathbf{x}_{n_{h}}^{\mathbf{\intercal }})$ of $n_{h}$
rows of covariate vectors $\mathbf{x}_{i(h)}^{\mathbf{\intercal }}$
respectively. Then a Bayesian 2-level model (HLM)\ can be represented by: 
\end{subequations}
\begin{subequations}
\label{Bayesian 2 level}
\begin{eqnarray}
y_{i(h)}\,|\,\mathbf{x}_{i(h)} &\sim &f(y\,|\,\mathbf{x}_{i(h)}),\text{ \ }%
i(h)=1,\ldots ,n_{h} \\
f(y\,|\,\mathbf{x}_{i(h)}) &=&\text{\textrm{n}}(y\,|\,\mathbf{x}%
_{i(h)}^{\intercal }\boldsymbol{\beta }_{Rh},\sigma ^{2}) \\
\mathbf{x}^{\intercal }\boldsymbol{\beta }_{Rh} &=&\mathbf{x}^{\intercal }%
\boldsymbol{\beta }+\mathbf{x}^{\intercal }\boldsymbol{u}_{h} \\
\beta _{0}\,|\,\sigma ^{2} &\sim &\text{$\mathrm{N}$}(0,\sigma ^{2}v_{\beta
0}\rightarrow \infty ) \\
\beta _{k}\,|\,\sigma ^{2} &\sim &\text{$\mathrm{N}$}(0,\sigma ^{2}v_{\beta
}),\text{ \ }k=1,\ldots ,p \\
\boldsymbol{u}_{h}\,|\,\boldsymbol{T} &\sim &\text{$\mathrm{N}$}(\mathbf{0},%
\boldsymbol{T}),\text{ \ }h=1,\ldots ,N_{h}  \label{NormMix} \\
\sigma ^{2} &\sim &\text{\textrm{IG}}(a_{0}/2,a_{0}/2) \\
\boldsymbol{T} &\sim &\text{\textrm{IW}}(p+3,s_{0}\mathbf{I}_{p+1}).
\end{eqnarray}%
This model (\ref{Bayesian 2 level}), as shown in (\ref{NormMix}), assumes
that the random coefficients $\boldsymbol{u}_{h}$ (for $h=1,\ldots ,N_{h}$)
are normally distributed over the $N_{h}$ groups.

Both linear models above, and the different versions of these models
mentioned in Section 1, are provided by the Bayesian Regression software.
See the Help menu for more details.

\section{Using the Bayesian Regression Software}

\subsection{Installing the\ Software}

The \textbf{Bayesian Regression} software is a stand-alone package for a
64-bit Windows computer\footnote{%
An older version of the software can run on a 32-bit computer.}. To install
the software on your computer, take the following steps:

\begin{enumerate}
\item Go the \textbf{Bayesian Regression} software web page:\newline
http://www.uic.edu/\symbol{126}georgek/HomePage/BayesSoftware.html.\newline
Then click the link on that page to download the \textbf{Bayesian Regression}
software installation file, named BayesInstaller\_web64bit.exe (or
BayesInstaller\_web32bit.exe).

\item Install the software by clicking the file
BayesInstaller\_webXXbit.exe. This will include a web-based installation of
MATLAB Compiler Runtime, if necessary. As you install, select the option
"Add a shortcut to the desktop," for convenience. (To install, be connected
to the internet, and temporarily disable any firewall or proxy settings on
your computer).
\end{enumerate}

\noindent Then start the software by clicking the icon BayesRegXXbit.exe.

The next subsection provides step-by-step instructions on how to use the 
\textbf{Bayesian regression} software to perform a Bayesian analysis of your
data set. The software provides several example data files, described under
the Help menu. You can create them by clicking the File menu option: "Create
Bayes Data Examples file folder." Click the File menu option again to import
and open an example data set from this folder. The next subsection
illustrates the software through the analysis of the example data set
PIRLS100.csv.

The \textbf{Bayesian Regression} software, using your menu-selected Bayesian
model, outputs the data analysis results into space- and comma-delimited
text files with time-stamped names, which can be viewed in free \textit{%
NotePad++}. The comma-delimited output files include the posterior samples\
(.MC1), model fit residual (*.RES), and the model specification (*.MODEL)
files. The software also outputs the results into graph (figure *.fig)
files, which can then be saved into a EPS\ (*.eps), bitmap (*.bmp), enhanced
metafile (*.emf), JPEG image (*.jpg), or portable document (*.pdf) file
format. Optionally you may graph or analyze a delimited text output file
after importing it into spreadsheet software (e.g., \textit{OpenOffice}) or
into the R software using the command line:\ \texttt{ImportedData =
read.csv(file.choose())}.

\subsection{Running the Software (12 Steps for Data Analysis)}

You can run the software for data analysis using any one of many Bayesian
models of your choice. A data analysis involves running the following 12
basic steps (required or optional).

In short, the 12 steps are as follows:\ \newline
(1)\ \ Import or open the data file (Required);\newline
(2)\ \ Compute basic descriptive statistics and plots of your data
(Optional);\newline
(3)\ \ Modify the data set (e.g., create variables) to set up your data
analysis model (Optional);\newline
(4)\ \ Specify a new Bayesian model for data analysis (Required);\newline
(5)\ \ Specify observation weights (Optional);\newline
(6)\ \ Specify the censored observations (Optional);\newline
(7)\ \ Set up the MCMC\ sampling algorithm model posterior estimation
(Required);\newline
(8)\ \ Click the \textbf{Run Posterior Analysis} button (Required);\newline
(9) \ Click the \textbf{Posterior Summaries} button to output data analysis
results (Required);\newline
(10)\ \ Check MCMC\ convergence (Required);\newline
(11)\ \ Click the \textbf{Posterior Predictive button} to run model
predictive analyses (Optional);\newline
(12)\ \ Click the \textbf{Clear button} to finish your data analysis project.%
\newline
Then you may run a different data analysis. Otherwise, you may then Exit the
software and return to the same data analysis project later, after
re-opening the software.

Below, we give more details on the 12 steps of data analysis.

\begin{enumerate}
\item \textbf{(Required)} \ Use the \underline{\textbf{File}}\textbf{\ menu}
to \textbf{Import or open the data file} for analysis. Specifically, the
data file that you import must be a comma-delimited file, with name having
the .csv extension. (Or, you may click the File menu option to open an
existing (comma-delimited) data (*.DAT)\ file). In the data file, the
variable names are located in the first row, with numeric data (i.e.,
non-text data) in all the other rows. For each row, the number of variable
names must equal the number of commas minus 1. The software allows for
missing data values, each coded as NaN or as an empty blank. After you
select the data file to import, the software converts it into a
comma-delimited data (*.DAT) file. Figure 1 shows the interface of the 
\textbf{Bayesian regression} software. It presents the PIRLS100.DAT\ data
set at the bottom of the interface, after the PIRLS100.csv file has been
imported.

\item \textbf{(Optional)} \ Use the \underline{\textbf{Describe/Plot Data Set%
}}\textbf{\ menu} \textbf{option(s)} to compute basic descriptive statistics
and plots of the data variables. Statistics and plots include the sample
mean, standard deviation, quantiles, frequency tables, cross-tabulations,
correlations, covariances, univariate or bivariate histograms\footnote{%
For the univariate histogram, the bin size ($h$) is defined by the
Freedman-Diaconis (1981\nocite{FreedmanDiaconis81}) rule, with $h=2($IQR)$%
n^{-1/3}$, where IQR is the interquartile range of the data, and $n$ is the
sample size. For the bivariate histogram, the automatic bin sizes are given
by $h_{k}=3.5\widehat{\sigma }_{k}n^{-1/4}$, where $\widehat{\sigma }_{k}$, $%
k=1,2$, is the sample standard deviation for the two variables (Scott, 1992%
\nocite{Scott92}).}, stem-and-leaf plots, univariate or bivariate kernel
density estimates\footnote{%
The univariate kernel density estimate assumes normal kernels, with
automatic bandwidth ($h$) given by the normal reference rule, defined by $%
h=1.06\widehat{\sigma }n^{-1/5}$, where $\widehat{\sigma }$ is the data
standard deviation, and $n$ is the sample size (Silverman, 1986\nocite%
{Silverman86}, p. 45). The bivariate kernel density estimate assumes normal
kernels, with automatic bandwidth determined by the equations in Botev et
al. (2010\nocite{Botev10}).}, quantile-quantile (Q-Q) plots, two- or
three-dimensional scatter plots, scatter plot matrices, (meta-analysis)\
funnel plots (Egger et al. 1997\nocite{Egger_etal_97}), box plots, and plots
of kernel regression estimates with automatic bandwidth selection\footnote{%
Kernel regression uses normal kernels, with an automatic choice of bandwidth
($h$) given by $\widehat{h}=\sqrt{\widehat{h}_{x}\widehat{h}_{y}}$, where $%
\widehat{h}_{x}=\mathrm{med}(\,|\,\mathbf{x}-\mathrm{med}(\mathbf{x}%
)\,|\,)/c $, and $\widehat{h}_{y}=\mathrm{med}(\,|\,\mathbf{y}-\mathrm{med}(%
\mathbf{y})\,|\,)/c,$ where $(\mathbf{x,y)}$ give the vectors of $X$ data
and $Y$ data (resp.), $\mathrm{med}(\cdot )$ is the median, $c=.6745\ast
(4/3/n)^{0.2}$, and $n$ is the sample size (Bowman \& Azzalini, 1997, p. 31%
\nocite{BowmanAzzalini97})}.

\item \textbf{(Optional)}\ \ Use the \underline{\textbf{Modify Data Set}} 
\textbf{menu option(s) }to set up a data analysis. The menu options allow
you to construct new variables, handle missing data, and/or to perform other
modifications of the data set. Then the new and/or modified variables can be
included in the Bayesian model that you select in Step 4. Figure 1 presents
the PIRLS100.DAT data at the bottom of the software interface, and shows the
data of the variables MALE, AGE, CLSIZE, ELL, TEXP4, EDLEVEL, ENROL, and
SCHSAFE in the last 8 data columns, respectively, after taking z-score
transformations and adding "Z:" to each variable name. Such transformations
are done with the menu option: Modify Data Set \TEXTsymbol{>} Simple
variable transformations \TEXTsymbol{>}\ Z score. Section 4.3 provides more
details about the available Modify Data Set menu options.\newline
----- \ FIGURE 1 \ in \texttt{http://www.uic.edu/\symbol{126}%
georgek/HomePage/Figures.pdf} \ -----

\item \textbf{(Required)} \ \textbf{Click the }\underline{\textbf{Specify
New Model}} button to \textbf{select a Bayesian model} for data analysis,
and then for the model select: the dependent variable; the covariate(s)
(predictor(s)) (if selected a regression model); the level-2 (and possibly
level-3) grouping variables (if a multi-level model); and the model's prior
distribution parameters. Figure 2 shows the software interface, after
selecting the Infinite homoscedastic probits regression model, along with
the dependent variable, covariates, and prior parameters.

\item \textbf{(Optional)}\ \ To weight each data observation (row)\
differently under your selected model, \textbf{click the \underline{\textbf{%
Observation Weights}} button} to select a variable containing the weights
(must be finite, positive, and non-missing). (This button is not available
for a binary or ordinal regression model). By default, the observation
weights are 1. For example, observation weights are used for meta-analysis
of data where each dependent variable observation $y_{i}$ represents a
study-reported effect size (e.g., a standardized mean difference in scores
between a treatment group and a control group, or a correlation coefficient
estimate). Each reported effect size $y_{i}$ has sampling variance $\widehat{%
\sigma }_{i}^{2}$, and observation weight $1/\widehat{\sigma }_{i}^{2}$ that
is proportional to the sample size for $y_{i}$. Details about the various
effect size measures, and their sampling variance formulas, are found in
meta-analysis textbooks (e.g., Cooper, et al. 2009\nocite%
{CooperHedgesValentine09}). Section 4.3 mentions a Modify Data Set menu
option that computes various effect size measures and corresponding
variances.\newline
----- \ FIGURE 2 \ in \texttt{http://www.uic.edu/\symbol{126}%
georgek/HomePage/Figures.pdf} \ -----

\item \textbf{(Optional)}\ \ \textbf{Click the \underline{\textbf{Censor
Indicators of Y}}\ button}, if the dependent variable consists of \textbf{%
censored observations} (not available for a binary or ordinal regression
model). Censored observations often appear in survival data, where the
dependent variable $Y$ represents the (e.g., log) survival time of a
patient. Formally, an observation, $y_{i}$, is \textit{censored} when it is
only known to take on a value from a known interval $[Y_{LBi},Y_{UBi}]$; is 
\textit{interval-censored} when $-\infty <Y_{LBi}<Y_{UBi}<\infty $; is 
\textit{right censored} when $-\infty <Y_{LBi}<Y_{UBi}\equiv \infty $; and 
\textit{left censored} when $-\infty \equiv Y_{LBi}<Y_{UBi}<\infty $ (e.g.,
Klein \&\ Moeschberger, 2010\nocite{KleinMoeschberger10}). After clicking
the \textbf{Censor Indicators of Y\ button}, select the two variables that
describe the (fixed)\ censoring lower-bounds ($LB$) and upper-bounds ($UB$)
of the dependent variable observations. Name these variables $LB$ and $UB$.
Then for each interval-censored observation $y_{i}$, its $LB_{i}$\ and $%
UB_{i}$ values must be finite, with $LB_{i}<UB_{i}$, $y_{i}\leq UB_{i}$, and 
$y_{i}\geq LB_{i}$. For each right-censored observation $y_{i}$, its $LB_{i}$%
\ value must be finite, with $y_{i}\geq LB_{i}$, and set $UB_{i}=-9999$. For
each left-censored observation $y_{i}$, its $UB_{i}$ value must be finite,
with $y_{i}\leq UB_{i}$, and set $LB_{i}=-9999$. For each uncensored
observation $y_{i}$, set $LB_{i}=-9999$ and $UB_{i}=-9999$.

\item \textbf{(Required)} \ \textbf{Enter: the total number (}$S$\textbf{)
of \underline{\textbf{MC\ Samples}}}, i.e., MCMC sampling iterations
(indexed by $s=1,\ldots ,S$, respectively)\textbf{; }the number ($s_{0}\geq
1 $) of the initial \underline{\textbf{Burn-In}} \textbf{period} samples;
and the \underline{\textbf{Thin}} number $k$, to retain every $k^{\text{th}}$
sampling iterate of the $S$ total MCMC samples. The MCMC\ samples are used
to estimate the posterior distribution (and functionals) of the parameters
of your selected Bayesian model. Entering a thin value $k>1$ represents an
effort to have the MCMC\ samples be (pseudo-) independent samples from the
posterior distribution. The burn-in number ($s_{0}$) is your estimate of the
number of initial MCMC\ samples that are biased by the software's starting
model parameter values used to initiate the MCMC chain at iteration $s=0$.

\item \textbf{(Required)} \ \textbf{Click the \underline{\textbf{Run
Posterior Analysis}} button} to run the MCMC\ sampling algorithm, using the
selected MC Samples, Burn-In, and Thin numbers. A wait-bar will then appear
and display the progress of the MCMC\ sampling iterations. After the MCMC\
sampling algorithm finishes running, the software will create an external:
(a)\ model (*.MODEL)\ text file that describes the selected model and data
set; (b)\ Monte Carlo (*.MC1) samples file which contains the generated
MCMC\ samples; (c)\ residual (*.RES)\ file that contains the model's
residual fit statistics; and (d)\ an opened, text output file of summaries
of (marginal)\ posterior point-estimates of model parameters and other
quantities, such as model predictive data-fit statistics. The results are
calculated from the generated MCMC samples aside from any burn-in or
thinned-out samples. Model fit statistics are calculated from all MCMC
samples instead. The software creates all output files in the same
subdirectory that contains the data (*.DAT) file.

\item \textbf{(Required)} \ Click the \underline{\textbf{Posterior Summaries}%
}\textbf{\ button} to select menu options for additional data analysis
output, such as: text output of posterior quantile estimates of model
parameters and 95\%\ Monte Carlo Confidence Intervals (see Step 10); trace
plots of MCMC\ samples; 2-dimensional plots and 3-dimensional plots of
(kernel)\ density estimates, univariate and bivariate histograms,
distribution function, quantile function, survival function, and hazard
functions, box plots, Love plots, Q-Q plots, and Wright maps, of the
(marginal) posterior distribution(s) of the model parameters; posterior
correlations and covariances of model parameters; and plots and tables of
the model's standardized fit residuals. The software creates all text output
files in the same subdirectory that contains the data (*.DAT) file. You may
save any graphical output in the same directory.

\item \textbf{(Required)} \ Click the \underline{\textbf{Posterior Summaries}%
} button for menu options to \textbf{check the MCMC\ convergence of
parameter point-estimates}, for every model parameter of interest for data
analysis. Verify:\ (1) that the univariate trace plots present good mixing
of the generated MCMC\ samples of each parameter; and (2)\ that the
generated MCMC\ samples of that parameter provide sufficiently-small
half-widths of the 95\%\ Monte Carlo Confidence Intervals (95\%\ MCCIs) for
parameter posterior point-estimates of interest (e.g., marginal posterior
mean, standard deviation, quantiles, etc.). If for your model parameters of
interest, either the trace plots do not support adequate mixing (i.e., plots
are not stable and "hairy"), or the 95\%\ MCCI half-widths are not
sufficiently small for practical purposes, then the MCMC samples of these
parameters have not converged to samples from the model's posterior
distribution. In this case you need to generate additional MCMC\ samples, by
clicking the \underline{\textbf{Run Posterior Analysis}} \textbf{button}
again. Then re-check for MCMC\ convergence by evaluating the updated trace
plots and the 95\% MCCI half-widths. This process may be repeated until
MCMC\ convergence is reached.

\item \textbf{(Optional)}\ \ \textbf{Click the \underline{\textbf{Posterior
Predictive}} button}\footnote{%
This button is not available for Bayesian models for density estimation.
However, for these models, the software provides menu options to output
estimates of the posterior predictive distribution, after clicking the 
\textbf{Posterior Summaries} button. They include density and c.d.f.
estimates. See Step 9.} to generate model's predictions of the dependent
variable $Y$, conditionally on selected values of one or more (focal)\
covariates (predictors). See Section 2 for more details.\ Then select the
posterior predictive functionals of $Y$ of interest.\ Choices of functionals
include the mean, variance, quantiles (to provide a quantile regression
analysis), probability density function (p.d.f.), cumulative distribution
function (c.d.f.), survival function, hazard function, the cumulative hazard
function, and the probability that $Y\geq 0$. Then select one or more focal
covariate(s) (to define $\mathbf{x}_{\mathcal{S}}$), and then enter their
values, in order to study how predictions of $Y$ varies as a function of
these covariate values. For example, if you chose the variable Z:CLSIZE as a
focal covariate, then you may enter values like $-1.1,$ $.02,$ $3.1,$ so
that you can make predictions of $Y$\ conditionally on these covariate
values. Or you may base predictions on an equally-spaced grid of covariate
(Z:CLSIZE) values, like $-3,-2.5,-2,\ldots 2,2.5,3$, by entering $-3:.5:3.$
If your data set observations are weighted (optional Step \#5), then specify
a weight value for the $Y$\ predictions. Next, if your selected focal
covariates do not constitute all model covariates, then select among options
to handle the remaining (non-focal) covariates. Options include the \textit{%
grand-mean centering method}, the \textit{zero-centering method}, the 
\textit{partial dependence method}, and the \textit{clustered partial
dependence method. }After you made all the selections, the software will
provide estimates of your selected posterior predictive functionals of $Y$,
conditionally on your selected covariate values, in graphical and text
output files, including comma-delimited files. (The software generates
graphs only if you selected 1 or 2 focal covariates; and generates no output
if you specify more than 300 distinct values of the focal covariate(s)). All
analysis output is generated in the same subdirectory that contains the data
(*.DAT)\ file. You may save any graphical output in the same directory.

\item \textbf{(Required)} \ After completing the Bayesian data analysis, you
may click the \underline{\textbf{Clear}}\textbf{\ button}. Then you may
start a different data analysis with another Bayesian model, or exit the
software. Later, you may return to and continue from a previous Bayesian
data analysis (involving the same model and data set) by using menu options
to generate new MCMC\ samples and/or new data analysis output. To return to
the previous analysis, go to the File menu to open the (relevant)\ data file
(if necessary), and then click the \underline{\textbf{Open\ Model}}\textbf{\
button} to open the old model (*.MODEL) file. Then, the software will load
this model file along with the associated MCMC samples (*.MC1) file and
residual (*.RES) files. (Returning to a previous Bayesian regression
analysis is convenient if you have already stored the data, model, MC
samples, and residual files all in the same file subdirectory). Then after
clicking the \underline{\textbf{Run Posterior Analysis}}\textbf{\ button},
the newly generated MCMC\ samples will append the existing MCMC samples
(*.MC1) file and update the residual (*.RES) file.
\end{enumerate}

\noindent Finally, the software provides a \textbf{Fast Ridge Regression}
menu option that performs a Bayesian data analysis using the ridge (linear)\
regression model (Hoerl \&\ Kennard, 1970\nocite{HoerlKennard70}), with
parameters estimated by a fast marginal maximum likelihood algorithm
(Karabatsos, 2014\nocite{Karabatsos14c}). This menu option can provide a
fast analysis of an ultra-large data set, involving either a very large
sample size and/or number of covariates (e.g., several thousands). At this
point we will not elaborate on this method because it is currently the
subject of ongoing research.

\subsection{Modify Data Set Menu Options}

Some \textbf{Modify Data Set} \textbf{menu} options allow you to construct
new variables from your data. These new variables may be included as either
covariates and/or the dependent variable for your Bayesian data analysis
model. Methods for constructing new variables include: simple
transformations of variables (e.g., z-score, log, sum of variables);
transforming a variable into an effect size dependent variable for
meta-analysis (Borenstein, 2009\nocite{Borenstein09}; Fleiss \&\ Berlin, 2009%
\nocite{FleissBerlin09}; Rosenthal, 1994\nocite{Rosenthal94}); the creation
of lagged dependent variables as covariates for a Bayesian autoregression
time-series analysis (e.g., Prado \&\ West, 2010\nocite{PradoWest2010});
dummy/binary\ coding of variables; the construction of new covariates from
other variables (covariates), via transformations including:\ polynomials,
two-way interactions between variables, univariate or multivariate
thin-plate splines (Green \&\ Silverman, 1993\nocite{GreenSilverman93}) or
cubic splines (e.g., Denison, et al. 2002\nocite{DenisonHolmesMallickSmith02}%
); spatial-weight covariates (Stroud, et al. 2001\nocite{StroudMullerSanso01}%
; or thin-plate splines; Nychka, 2000\nocite{Nychka00}) from\ spatial data
(e.g., from latitude and longitude variables) for spatial data analysis.

Now we briefly discuss Modify Data Set menu options that can help set up a
causal analysis of data from a non-randomized (or randomized) study. First,
a propensity score variable, included as a covariate in a regression model;
or as a dummy-coded covariate that stratifies each subject into one of 10
(or more)\ ordered groups of propensity scores; or as observations weights
(entered as the inverse of the propensity scores); can help reduce selection
bias in the estimation of the causal effect (slope coefficient)\ of a
treatment-receipt (versus non-treatment/control) indicator covariate on a
dependent variable of interest (Rosenbaum \&\ Rubin, 1983\nocite%
{RosenbaumRubin83a}, 1984\nocite{RosenbaumRubin84}; Imbens, 2004\nocite%
{Imbens04}; Lunceford \&\ Davidian, 2004\nocite{LuncefordDavidian04};
Schafer \&\ Kang, 2008\nocite{SchaferKang08}; Hansen, 2008\nocite{Hansen08}%
). As an alternative to using propensity scores, we may consider a
regression discontinuity design analysis (Thistlewaite \&\ Campbell, 1960%
\nocite{ThistlewaiteCampbell60}; Cook, 2008\nocite{Cook08}). This would
involve specifying a regression model, with dependent (outcome)\ variable of
interest regressed on covariates that include an assignment variable, and a
treatment assignment variable that indicates (0 or 1)\ whether or not the
assignment variable exceeds a meaningful threshold. Then under mild
conditions (Hahn, et al. 2001\nocite{HahnToddVanderKlaauw01}; Lee \&\
Lemieux, 2010\nocite{LeeLemieux10}), the slope coefficient estimate for the
treatment variable is a causal effect estimate of the treatment (versus
non-treatment)\ on the dependent variable. For either the propensity score
or regression discontinuity design approach, which can be set up using
appropriate Modify Data Set menu options, causal effects can be expressed in
terms of treatment versus non-treatment comparisons of general posterior
predictive functionals of the dependent variable (e.g., Karabatsos \&\
Walker, 2015\nocite{KarabatsosWalker15}).

In some settings, it is of interest to include a multivariate dependent
variable (multiple dependent variables) in the Bayesian regression model
(Step 4). You can convert a multivariate regression problem into a
univariate regression problem (Gelman, et al., 2004, Ch. 19\nocite%
{GelmanCarlinSternRubin04}) by clicking a menu option that "vectorizes" or
collapses the multiple dependent variables into a single dependent variable.
This also generates new covariates in the data set, having a block design
that is suited for the (vectorized) multiple dependent variables. To given
an example involving item response theory (IRT)\ modeling (e.g., van der
Linden, 2015\nocite{vanderLinden15}), each examinee (data set row) provides
data on responses to $J$ items (data columns) on a test (e.g., examination
or questionnaire). Here, a menu option can be used to vectorize (collapse)\
the $J$ item response columns into a single new dependent variable (data
column), and to create corresponding dummy item indicator ($-1$ or $0$)
covariates. In the newly-revised data, each examinee occupies $J$\ rows of
data. Then a Bayesian regression model of the software, which includes the
new dependent variable, item covariates, and the examinee identifier as a
grouping variable (or as dummy (0,1)\ indicator covariates), provides an
Item Response Theory (IRT) model for the data (van der Linden, 2015\nocite%
{vanderLinden15}).

Similarly, it may be of interest to include a categorical (polychotomous)
dependent variable in your Bayesian regression model. Assume that the
variable takes on $m+1$ (possibly unordered) categorical values, indexed by $%
c=0,1,\ldots ,m$, with $c=0$ the reference category. In this case, you may
run a menu option that recodes the categorical dependent variables into $m+1$
binary (e.g., 0 or 1) variables, then vectorizes (collapses)\ these multiple
binary variables (data columns)\ into a single binary\ dependent variable
(column), and reformats the covariate data into a block design suitable for
the multiple binary variables. Then for data analysis, you may specify a
binary regression model that includes the (collapsed) binary dependent
variable and the reformatted covariates (Begg \&\ Gray, 1984\nocite%
{BeggGray1984}; also see Wu et al. 2004\nocite{WuLinWeng04}).

The Modify Data Set menu also provides menu options to reduce the number of
variables for dimension reduction, including: principal components,
multidimensional scaling, or scaling by the true score test theory and
Cronbach's alpha reliability analysis of test items, and propensity scoring.
There are also menu options that handle missing data, including:
nearest-neighbor hot-deck imputation of missing data (Andridge \&\ Little,
2010\nocite{AndridgeLittle10}); the processing of multiple (missing data)\
imputations or plausible values obtained externally; and the assignment of
missing (NaN)\ values based on specific missing data codes\ (e.g., 8, 9, or
-99, etc.)\ of variables. Finally, there are Modify Data Set menu options
for more basic data editing and reformatting, including: adding data on a
row-identification (ID) variable; aggregating (e.g., averaging)\ variables
by a group identification variable; the selection of cases (data rows) for
inclusion in a reduced data set, including random selection or selection by
values of a variable; the sorting of cases; the deletion of variables;
changing the variable name; and moving the variables into other columns of
the data file.\newline

\section{Real Data Example}

We illustrate the software through the analysis of data set file
PIRLS100.csv, using the software steps outlined in the previous section. The
data are based on a random sample of 100 students from 21 low-income U.S.
elementary schools, who took part of the 2006 Progress in International
Reading Literacy Study (PIRLS).

\begin{center}
----- \ FIGURE 3 \ in \texttt{http://www.uic.edu/\symbol{126}%
georgek/HomePage/Figures.pdf} \ -----
\end{center}

The dependent variable is student literacy score (zREAD).\ The eight
covariates are: student male status (Z:MALE) indicator (0 or 1) and age
(AGE); student's class size (SIZE) and class percent English language
learners (ELL); student's teacher years of experience (TEXP4) and education
level (TEXP4 = 5 if bachelor's degree; TEXP4 = 6 if at least master's
degree); student's school enrollment (ENROL) and safety rating (SAFE = 1 is
high; SAFE = 3 is low).\noindent \noindent \noindent\ Figure 3 presents the
univariate descriptive statistics of these variables, from a text output
file obtained by the menu option: Describe/Plot Data Set\textbf{\ 
\TEXTsymbol{>}} Summaries and Frequencies \TEXTsymbol{>} Univariate
descriptives. Data is also available on a school identifier variable (SCHID).

The data for each of the 8 covariates were transformed into z-scores having
mean 0 and variance 1, using the menu option: Modify Data Set \TEXTsymbol{>}%
\ Simple variable transformations \TEXTsymbol{>}\ Z-score. These
transformations will make the slope coefficients of the 8 covariates\
(resp.)\ interpretable on a common scale, in a regression analysis.

The BNP\ infinite-probits mixture model (\ref{IP mixture reg model}) was fit
to the PIRLS100 data, using the dependent variable and the 8 z-scored
covariates. For the model's prior density function (\ref{IP prior pdf}), the
following software defaults were specified for the prior parameters: $%
b_{\sigma \mu }=5$, $v=100$, $a_{0}=0.01$, $v_{\omega }=10$, and $a_{\omega
}=0.01$. This represents an attempt to specify a rather non-informative
prior distribution for the model parameters. In order to estimate the
model's posterior distribution, 50,000 MCMC sampling iterations were run,
using an initial burn-in of 2,000 and thinning interval of 5.

Figure 4 presents the model's predictive data-fit statistics (including
their 95\%\ MCCI\ half-widths), from text output that was opened by the
software immediately after the 50K MCMC sampling iterations were run. The
model obtained a $D(\underline{m})$ statistic of $32$, with an R-squared of $%
.96$, and had no outliers according to standardized residuals that ranged
within $-2$ and $2$.

\begin{center}
\noindent ----- \ FIGURE 4 \ in \texttt{http://www.uic.edu/\symbol{126}%
georgek/HomePage/Figures.pdf} \ -----

----- \ FIGURE 5 \ in \texttt{http://www.uic.edu/\symbol{126}%
georgek/HomePage/Figures.pdf} \ -----
\end{center}

Figure 5 presents the (marginal) posterior point-estimates of all the model
parameters, calculated from the 50K\ MCMC\ samples (aside from the burn-in
and thinned-out MCMC\ samples).\ Estimates include the marginal posterior
mean, median, standard deviation, 50\%\ credible interval (given by the 25\%
(.25 quantile)\ and 75\%\ (.75 quantile)\ output columns), and the 95\%\
credible interval (2.5\%, 97.5\%\ columns). Figure 5 is part of the same
text output that reported the model fit statistics (Figure 4).

Figure 6 is a box plot of the (marginal) posterior quantile point-estimates
of the intercept and slope coefficient parameters for the 8 covariates
(i.e., parameters $\boldsymbol{\beta }=(\beta _{0},\beta _{1},\ldots ,\beta
_{p})^{\intercal }$). A red box (blue box, resp.)\ flags a coefficient
parameter that is (not, resp.)\ significantly different than zero, according
to whether or not the 50\%\ (marginal) posterior interval (box) includes
zero. The box plot menu option is available after clicking the Posterior
Summaries button.

\begin{center}
----- \ FIGURE 6 \ in \texttt{http://www.uic.edu/\symbol{126}%
georgek/HomePage/Figures.pdf} \ -----
\end{center}

MCMC convergence, for the parameters of interest for data analysis, can be
evaluated by clicking certain menu options after clicking the Posterior
Summaries button. Figure 7 presents the trace plots for the model's
intercept and slope parameters sampled over the 50K MCMC sampling
iterations. The trace plots appear to support good mixing for these
parameters, as each trace plot appears stable and "hairy." Figure 8 presents
the 95\%\ MCCI half-widths of the (marginal) posterior coefficient
point-estimates of Figure 5. The half-widths are nearly all less than .10,
and thus these posterior point-estimates are reasonably accurate in terms of
Monte Carlo standard error. If necessary, the software can re-click the
Posterior Analysis button, to run additional MCMC sampling iterations, to
obtain more precise posterior point-estimates of model parameters (as would
be indicated by smaller 95\%\ MCCI half-widths).

\begin{center}
----- \ FIGURE 7 \ in \texttt{http://www.uic.edu/\symbol{126}%
georgek/HomePage/Figures.pdf} \ -----

----- \ FIGURE 8 \ in \texttt{http://www.uic.edu/\symbol{126}%
georgek/HomePage/Figures.pdf} \ -----

----- \ FIGURE 9 \ in \texttt{http://www.uic.edu/\symbol{126}%
georgek/HomePage/Figures.pdf} \ -----

----- \ FIGURE 10 \ in \texttt{http://www.uic.edu/\symbol{126}%
georgek/HomePage/Figures.pdf} \ -----
\end{center}

The user can click the Posterior Predictive button to investigate how chosen
features (functionals)\ of the posterior predictive distribution varies as a
function of one or more selected (focal)\ covariates, through graphical and
text output.

Figure 9\ provides a quantile and mean regression analysis, by showing the
estimates of the mean, and the .1, .25, .5 (median), .75, and .9 quantiles
of the model's posterior predictive distribution of zREAD, conditionally on
selected values $-2,-1.5,\ldots ,1,1.5,2$ of the Z:CLSIZE covariate, and on
zero for all the 7 other covariates (using the zero-centering method). It
presents nonlinear relationships between literacy performance (zREAD) and
class size (Z:CLSIZE). In particular, for higher literacy-performing
students (i.e., .75 and .90 quantiles of zREAD), there is a highly nonlinear
relationship between zREAD\ and class size. For lower performing students
(i.e., .1 and .25 quantiles of zREAD), there is a flatter relationship
between zREAD\ and class size. For "middle" performing students (mean and .5
quantiles of zREAD), there is a rather flat relationship between zREAD\ and
class size.

Figure 10 shows a 3-dimensional plot of (Rao-Blackwellized) estimates of the
model's posterior predictive p.d.f. of zREAD, conditionally on the same
values of the 8 covariates mentioned earlier. As shown, both the location
and shape of the zREAD distribution changes as a function of class size. For
each posterior predictive p.d.f. estimate (panel), a homogeneous cluster of
students (in terms of a shared zREAD score)\ is indicated by a bump (or
mode)\ in the p.d.f.

\begin{center}
----- \ FIGURE 11 \ in \texttt{http://www.uic.edu/\symbol{126}%
georgek/HomePage/Figures.pdf} \ -----
\end{center}

The ANOVA-linear DDP model (\ref{ANOVA DDP}) was fit to the PIRLS100 data.
For this model's prior (\ref{ANOVA DDP prior pdf}), the following parameters
were chosen: $r_{0}=10$, $s_{0}=10$, $a_{0}=2,$ $a_{\alpha }=1,$ and $%
b_{\alpha }=1$. This was an attempt to specify a rather non-informative
prior distribution for the model parameters. This model was estimated using
50,000 MCMC sampling iterations, with burn-in of 2,000 and thinning interval
of 5. MCMC convergence was confirmed according to univariate trace plots and
small 95\% MCCI half-widths for (marginal)\ posterior point-estimates
(nearly all less than .1). Figure 11 presents the marginal posterior
densities (distributions) of the intercept and slope parameters, from the
mixture distribution of the model. Each of these distributions (densities)\
are skewed and/or multimodal, unlike normal distributions. Each mode (bump)\
in a distribution (density)\ refers to a homogeneous cluster of schools, in
terms of a common value of the given (random) coefficient parameter.

Finally, recall that for the PIRLS100 data, the infinite-probits mixture
model (\ref{IP mixture reg model}) obtained a $D(\underline{m})$ statistic
of $32$, an R-squared of $.96$, and no outliers according to all the
absolute standardized residuals being less than 2 (Figure 4). The
ANOVA-linear DDP\ model (\ref{ANOVA DDP}) obtained a $D(\underline{m})$
statistic of $113$, an R-squared of $.64$, and no outliers. The ordinary
linear model obtained a $D(\underline{m})$ statistic of $139$, an R-squared
of $.24$ and $5$ outliers.\ This linear model assumed rather non-informative
priors, with $\beta _{0}\sim \mathrm{N}(0,v_{0}\rightarrow \infty )$, $\beta
_{k}\sim \mathrm{N}(0,1000)$, $k=1,\ldots ,8$, and $\sigma ^{2}\sim $ 
\textrm{IG}$(.001,.001)$.

\section{Conclusions}

We described and illustrated a stand-alone, user-friendly and menu-driven
software package for Bayesian data analysis. Such analysis can be performed
using any one of many Bayesian models that are available from the package,
including BNP\ infinite-mixture regression models and normal random-effects
linear models. As mentioned in the data analysis exercises of Appendix B,
the software can be used to address a wide range of statistical applications
that arise from many scientific fields. In the future, new Bayesian mixture
models will be added to the software. They will include mixture models
defined by other kernel density functions, and models with parameters
assigned a mixture of P\'{o}lya Trees prior.

\section{Acknowledgments}

This paper is supported by NSF-MMS research grant SES-1156372.

\newpage

\begin{center}
{\Large References}
\end{center}

\begin{description}
\item Agresti A (2002). \textit{Categorical Data Analysis}. John Wiley and
Sons.

\item Albert J, Chib S (1993). \textquotedblleft Bayesian Analysis of Binary
and Polychotomous Response Data.\textquotedblright\ \textit{Journal of the
American Statistical Association}, \textit{88}, 669--679.

\item Andridge RR, Little R (2010). \textquotedblleft A Review of Hot Deck
Imputation for Survey Non-response.\textquotedblright\ \textit{International
Statistical Review}, \textit{78}, 40--64.

\item Antoniak C (1974). \textquotedblleft Mixtures of Dirichlet Processes
with Applications to Bayesian Nonparametric Problems.\textquotedblright\ 
\textit{Annals of Statistics}, \textit{2}, 1152--1174.

\item Atchad\'{e} Y, Rosenthal J (2005). \textquotedblleft On Adaptive
Markov chain Monte Carlo Algorithms.\textquotedblright\ \textit{Bernoulli}, 
\textit{11}, 815--828.

\item Barry D, Hartigan J (1993). \textquotedblleft A Bayesian-Analysis for
Change Point Problems.\textquotedblright\ \textit{Journal of the American
Statistical Association}, \textit{88}, 309--319.

\item Begg C, Gray R (1996). \textquotedblleft Calculation of Polychotomous
Logistic Regression Parameters Using Individualized
Regressions.\textquotedblright\ \textit{Biometrika}, \textit{71}, 11--18.

\item Bernardo J, Smith A (1994). \textit{Bayesian Theory}. Wiley,
Chichester, England.

\item Bertoin J (1998). \textit{L\'{e}vy Processes}. Cambridge University
Press, Cambridge.

\item Borenstein M (2009). \textquotedblleft Effect Sizes for Continuous
Data.\textquotedblright In H Cooper, L Hedges, J Valentine (eds.), \textit{%
The} \textit{Handbook of Research Synthesis and Meta-Analysis} (2nd Ed.),
pp. 221--235. Russell Sage Foundation, New York.

\item Botev Z, Grotowski J, Kroese D (2010). \textquotedblleft Kernel
Density Estimation via Diffusion.\textquotedblright\ \textit{The Annals of
Statistics}, \textit{38}, 2916--2957.

\item Bowman A, Azzalini A (1997). \textit{Applied Smoothing Techniques for
Data Analysis: The Kernel Approach with S-Plus Illustrations}. Oxford
University Press, Oxford.

\item Brooks S (1998). \textquotedblleft Quantitative Convergence Assessment
for Markov chain Monte Carlo via Cusums.\textquotedblright\ \textit{%
Statistics and Computing,} \textit{8}, 267--274.

\item Brooks S, Gelman A, Jones G, Meng X (2011). \textit{Handbook of Markov
Chain Monte Carlo}. Chapman and Hall/CRC, Boca Raton, FL.

\item Burr D (2012). \textquotedblleft bspmma: An R package for Bayesian
Semiparametric Models for Meta-Analysis.\textquotedblright\ \textit{Journal
of Statistical Software}, \textit{50}, 1--23.

\item Burr D, Doss H (2005). \textquotedblleft A Bayesian Semiparametric
Model for Random-Effects Meta-Analysis.\textquotedblright\ \textit{Journal
of the American Statistical Association}, \textit{100}, 242--251.

\item Casella G, Berger R (2002). \textit{Statistical Inference (2nd Ed.)}.
Duxbury, Pacific Grove, CA.

\item Cepeda E, Gamerman D (2001). \textquotedblleft Bayesian modeling of
variance heterogeneity in normal regression models.\textquotedblright\ 
\textit{Brazilian Journal of Probability and Statistics}, \textit{14},
207--221.

\item Cook T (2008). \textquotedblleft Waiting for Life to Arrive: A History
of the Regression-Discontinuity Design in Psychology, Statistics and
Economics.\textquotedblright\ \textit{Journal of Econometrics}, \textit{142}%
, 636--654.

\item Cooper H, Hedges L, Valentine J (2009). \textit{The Handbook of
Research Synthesis and Meta-Analysis (Second Edition)}. Russell Sage
Foundation, New York.

\item De Iorio M, M\"{u}ller P, Rosner G, MacEachern S (2004).
\textquotedblleft An ANOVA model for dependent random
measures.\textquotedblright\ \textit{Journal of the American Statistical
Association}, \textit{99}, 205--215.

\item DeBlasi P, Favaro S, Lijoi A, Mena R, Pr\"{u}nster I, Ruggiero M
(2015). \textquotedblleft Are Gibbs-Type Priors the Most Natural
Generalization of the Dirichlet Process?\textquotedblright\ \textit{IEEE
Transactions on Pattern Analysis and Machine Intelligence}, \textit{37},
212--229.

\item Denison D, Holmes C, Mallick B, Smith A (2002). \textit{Bayesian
Methods for Nonlinear Classification and Regression}. John Wiley and Sons,
New York.

\item Di Lucca M, Guglielmi A, M\"{u}ller P, Quintana F (2012).
\textquotedblleft A Simple Class of Bayesian Non-parametric Autoregression
Models.\textquotedblright \textit{\ Bayesian Analysis}, \textit{7, }771--796.

\item Egger M, Smith GD, Schneider M, Minder C (1997). \textquotedblleft
Bias in Meta-Analysis Detected by a Simple, Graphical
Test.\textquotedblright\ \textit{British Medical Journal}, \textit{315},
629--634.

\item Escobar M, West M (1995). \textquotedblleft Bayesian Density
Estimation and Inference Using Mixtures.\textquotedblright\ \textit{Journal
of the American Statistical Association}, \textit{90}, 577--588.

\item Evans I (1965). \textquotedblleft Bayesian Estimation of Parameters of
a Multivariate Normal Distribution.\textquotedblright\ \textit{Journal of
the Royal Statistical Society}, \textit{Series B}, \textit{27}, 279--283.

\item Favaro S, Lijoi A, Pr\"{u}nster I (2012). \textquotedblleft On the
Stick-Breaking Representation of Normalized Inverse Gaussian
Priors.\textquotedblright\ \textit{Biometrika}, \textit{99}, 663--674.

\item Ferguson T (1973). \textquotedblleft A Bayesian Analysis of Some
Nonparametric Problems.\textquotedblright\ \textit{Annals of Statistics}, 
\textit{1}, 209--230.

\item Flegal J, Jones G (2011). \textquotedblleft Implementing Markov Chain
Monte Carlo: Estimating with Confidence.\textquotedblright\ In S Brooks, A
Gelman, G Jones, X Meng (eds.), \textit{Handbook of Markov Chain Monte Carlo}%
, pp. 175--197. CRC, Boca Raton, FL.

\item Fleiss J, Berlin J (2009). \textquotedblleft Effect Size for
Dichotomous Data.\textquotedblright\ In H Cooper, L Hedges, J Valentine
(eds.), The \textit{Handbook of Research Synthesis and Meta-Analysis} (2nd
Ed.), pp. 237--253. Russell Sage Foundation, New York.

\item Freedman D, Diaconis P (1981). \textquotedblleft On the Histogram as a
Density Estimator: L2 theory.\textquotedblright\ P\textit{robability Theory
and Related Fields}, \textit{57}, 453--476.

\item Friedman J (2001). \textquotedblleft Greedy Function Approximation: A
Gradient Boosting Machine.\textquotedblright\ \textit{Annals of Statistics}, 
\textit{29}, 1189--1232.

\item Fuentes-Garc\'{\i}a R, Mena R, Walker S (2009). \textquotedblleft A
Nonparametric Dependent Process for Bayesian Regression.\textquotedblright\ 
\textit{Statistics and Probability Letters}, \textit{79}, 1112--1119.

\item Fuentes-Garc\'{\i}a R, Mena R, Walker S (2010). \textquotedblleft A
New Bayesian Nonparametric Mixture Model.\textquotedblright\ \textit{%
Communications in Statistics}, \textit{39}, 669--682.

\item Gelfand A, Diggle P, Guttorp P, Fuentes M (2010). \textit{Handbook of
Spatial Statistics}. Chapman and Hall/CRC, Boca Raton.

\item Gelfand A, Ghosh J (1998). \textquotedblleft Model Choice: A Minimum
Posterior Predictive Loss Approach.\textquotedblright\ \textit{Biometrika}, 
\textit{85}, 1--11.

\item Gelfand A, Mukhopadhyay S (1995). \textquotedblleft On Nonparametric
Bayesian Inference for the Distribution of a Random
Sample.\textquotedblright\ \textit{Canadian Journal of Statistics}, \textit{%
23}, 411--420.

\item Gelfand A, Smith A, Lee TM (1992). \textquotedblleft Bayesian Analysis
of Constrained Parameter and Truncated Data Problems Using Gibbs
Sampling.\textquotedblright\ \textit{Journal of the American Statistical
Association}, \textit{87}, 523--532.

\item Gelman A, Carlin A, Stern H, Rubin D (2004). \textit{Bayesian Data
Analysis (Second Edition)}. Chapman and Hall, Boca Raton, Florida.

\item George E, McCulloch R (1993). \textquotedblleft Variable Selection via
Gibbs Sampling.\textquotedblright\ \textit{Journal of the American
Statistical Association}, \textit{88}, 881--889.

\item George E, McCulloch R (1997). \textquotedblleft Approaches for
Bayesian Variable Selection.\textquotedblright\ \textit{Statistica Sinica}, 
\textit{7}, 339--373.

\item Geyer C (2011). \textquotedblleft Introduction to
MCMC.\textquotedblright In S Brooks, A Gelman, G Jones, X Meng (eds.), 
\textit{Handbook of Markov Chain Monte Carlo}, pp. 3--48. CRC, Boca Raton,
FL.

\item Ghosh J, Ramamoorthi R (2003). \textit{Bayesian Nonparametrics}.
Springer, New York.

\item Gilks W, Wang C, Yvonnet B, Coursaget P (1993). \textquotedblleft
Random-Effects Models for Longitudinal Data Using Gibbs
Sampling.\textquotedblright\ \textit{Biometrics}, \textit{49}, 441--453.

\item Green P, Silverman B (1993). \textit{Nonparametric Regression and
Generalized Linear Models: A Roughness Penalty Approach}. CRC Press, Los
Altos, CA.

\item Hahn J, Todd P, der Klaauw WV (2001). \textquotedblleft Identification
and Estimation of Treatment Effects with a Regression-Discontinuity
Design.\textquotedblright\ \textit{Econometrica}, \textit{69}, 201--209.

\item Hansen B (2008). \textquotedblleft The Prognostic Analogue of the
Propensity Score.\textquotedblright\ \textit{Biometrika}, \textit{95},
481--488.

\item Hanson T (2006). \textquotedblleft Inference for Mixtures of Finite P%
\'{o}lya Tree Models.\textquotedblright\ \textit{Journal of the American
Statistical Association}, \textit{101}, 1548--1565.

\item Hartigan J (1990). \textquotedblleft Partition
Models.\textquotedblright\ \textit{Communications in Statistics: Theory and
Methods}, \textit{19}, 2745--2756.

\item Hastie T, Tibshiriani R, Friedman J (2009). \textit{The Elements of
Statistical Learning: Data Mining, Inference, and Prediction (2nd Ed.)}.
Springer-Verlag, New York.

\item Hedges L (1981). \textquotedblleft Distribution Theory for Glass's
Estimator of Effect size and Related Estimators.\textquotedblright\ \textit{%
Journal of Educational and Behavioral Statistics}, \textit{6}, 107--128.

\item Hjort N, Holmes C, M\"{u}ller P, Walker S (2010). \textit{Bayesian
Nonparametrics}. Cambridge University Press, Cambridge.

\item Hoerl A, Kennard R (1970). \textquotedblleft Ridge Regression: Biased
Estimation for Nonorthogonal Problems.\textquotedblright\ \textit{%
Technometrics}, \textit{12}, 55--67.

\item Holmes C, Held K (2006). \textquotedblleft Bayesian Auxiliary Variable
Models for Binary and Multinomial Regression.\textquotedblright\ \textit{%
Bayesian Analysis}, \textit{1}, 145--168.

\item IBM Corp (2015). \textit{IBM SPSS Statistics for Windows, v22.0}. IBM
Corp., Armonk, NY.

\item Imbens G (2004). \textquotedblleft Nonparametric Estimation of Average
Treatment Effects Under Exogeneity: A Review.\textquotedblright\ \textit{The
Review of Economics and Statistics}, \textit{86}, 4--29.

\item Imbens GW, Lemieux T (2008). \textquotedblleft Regression
Discontinuity Designs: A Guide to Practice.\textquotedblright\ \textit{%
Journal of Econometrics}, \textit{142}, 615--635.

\item Ishwaran H, James L (2001). \textquotedblleft Gibbs Sampling Methods
for Stick-Breaking Priors.\textquotedblright\ \textit{Journal of the
American Statistical Association}, \textit{96}, 161--173.

\item Ishwaran H, Zarepour M (2000). \textquotedblleft Markov chain Monte
Carlo in Approximate Dirichlet and Beta Two-Parameter Process Hierarchical
Models.\textquotedblright\ \textit{Biometrika}, \textit{87}, 371--390.

\item James L, Lijoi A, Pr\"{u}nster I (2009). \textquotedblleft Posterior
Analysis for Normalized Random Measures with Independent
Increments.\textquotedblright \textit{\ Scandinavian Journal of Statistics}, 
\textit{36}, 76--97.

\item Jara A, Hanson T, Quintana F, M\"{u}ller P, Rosner G (2011).
\textquotedblleft DPpackage: Bayesian Semi- and Nonparametric Modeling in
R.\textquotedblright\ \textit{Journal of Statistical Software}, \textit{40},
1--20.

\item Jara A, Lesaffre E, DeIorio M, Quintana F (2010). \textquotedblleft
Bayesian Semiparametric Inference for Multivariate Doubly-Interval-Censored
Data.\textquotedblright\ \textit{Annals of Applied Statistics}, \textit{4},
2126--2149.

\item Johnson N, Kotz S, Balakrishnan N (1994). \textit{Continuous
Univariate Distributions, Vol. 1}. Wiley, New York.

\item Jordan M, Jacobs R (1994). \textquotedblleft Hierarchical Mixtures of
Experts and the EM Algorithm.\textquotedblright\ \textit{Neural Computation}%
, \textit{6}, 181--214.

\item Kalli M, Griffin J, Walker S (2011). \textquotedblleft Slice Sampling
Mixture Models.\textquotedblright\ \textit{Statistics and Computing}, 
\textit{21}, 93--105.

\item Karabatsos G (2014). \textquotedblleft Fast Marginal Likelihood
Estimation of the Ridge Parameter in Ridge Regression.\textquotedblright\
Technical Report 1409.2437, arXiv preprint.

\item Karabatsos G, Talbott E, Walker S (2015). \textquotedblleft A Bayesian
Nonparametric Meta-Analysis Model.\textquotedblright\ \textit{Research
Synthesis Methods}, \textit{6}, 28--44.

\item Karabatsos G, Walker S (2012a). \textquotedblleft Adaptive-Modal
Bayesian Nonparametric Regression.\textquotedblright\ \textit{Electronic
Journal of Statistics}, \textit{6}, 2038--2068.

\item Karabatsos G, Walker S (2012b). \textquotedblleft Bayesian
Nonparametric Mixed Random Utility Models.\textquotedblright\ \textit{%
Computational Statistics and Data Analysis}, \textit{56}, 1714--1722.

\item Karabatsos G, Walker S (2015). \textquotedblleft Bayesian
Nonparametric Item Response Models.\textquotedblright\ In W. van der Linden
(ed.), \textit{Handbook Of Item Response Theory, Volume 1: Models,
Statistical Tools, and Applications}. Taylor and Francis, New York.

\item Karabatsos G, Walker S (2015, to appear). \textquotedblleft A Bayesian
Nonparametric Causal Model for Regression Discontinuity
Designs.\textquotedblright\ In P M\"{u}ller, R Mitra (eds.), \textit{%
Nonparametric Bayesian Methods in Biostatistics and Bioinformatics}.
Springer-Verlag, New York. (ArXiv preprint 1311.4482).

\item Kingman J (1975). \textquotedblleft Random Discrete
Distributions.\textquotedblright\ \textit{Journal of the Royal Statistical
Society}, \textit{Series B}, \textit{37}, 1--22.

\item Klein J, Moeschberger M (2010). \textit{Survival Analysis (2nd Ed.)}.
Springer-Verlag, New York.

\item Laud P, Ibrahim J (1995). \textquotedblleft Predictive Model
Selection.\textquotedblright\ \textit{Journal of the Royal Statistical
Society}, \textit{Series B}, \textit{57}, 247--262.

\item Lee D (2008). \textquotedblleft Randomized Experiments from Non-Random
Selection in U.S. House Elections.\textquotedblright\ Journal of
Econometrics, 142, 675--697.

\item Lee D, Lemieux T (2010). \textquotedblleft Regression Discontinuity
Designs in Economics.\textquotedblright\ \textit{The Journal of Economic
Literature}, \textit{48}, 281--355.

\item Lijoi A, Mena R, Pr\"{u}nster I (2005). \textquotedblleft Hierarchical
Mixture Modeling with Normalized Inverse-Gaussian Priors.\textquotedblright\ 
\textit{Journal of the American Statistical Association}, \textit{100},
1278--1291.

\item Lindley D, Smith A (1972). \textquotedblleft Bayes Estimates for the
Linear Model (with Discussion).\textquotedblright\ \textit{Journal of the
Royal Statistical Society}, \textit{Series B}, \textit{34}, 1--41.

\item Lo A (1984). \textquotedblleft On a Class of Bayesian Nonparametric
Estimates.\textquotedblright\ \textit{Annals of Statistics}, \textit{12},
351--357.

\item Lunceford J, Davidian M (2004). \textquotedblleft Strati\ldots cation
and Weighting Via the Propensity Score in Estimation of Causal Treatment
Effects: A Comparative Study.\textquotedblright\ \textit{Statistics in
Medicine}, \textit{23}, 2937--2960.

\item MacEachern S (1999). \textquotedblleft Dependent Nonparametric
Processes.\textquotedblright\ \textit{Proceedings of the Bayesian
Statistical Sciences Section of the American Statistical Association}, pp.
50--55.

\item MacEachern S (2000). \textquotedblleft Dependent Dirichlet
Processes.\textquotedblright\ Technical report, Department of Statistics,
The Ohio State University.

\item MacEachern S (2001). \textquotedblleft Decision Theoretic Aspects of
Dependent Nonparametric Processes.\textquotedblright\ In E George (ed.), 
\textit{Bayesian Methods with Applications to Science, Policy and Official
Statistics}, pp. 551--560. International Society for Bayesian Analysis,
Creta.

\item McCullagh P, Nelder J (1989). \textit{Generalized Linear Models (2nd
Ed.)}. Chapman and Hall, London.

\item McLachlan G, Peel D (2000). \textit{Finite Mixture Models}. John Wiley
and Sons, New York.

\item Mitra R, M\"{u}ller P (2015, to appear). \textit{Nonparametric
Bayesian Methods in Biostatistics and Bioinformatics}. Springer-Verlag, New
York.

\item Molenberghs G, Verbeke G (2005). \textit{Models for Discrete
Longitudinal Data}. Springer-Verlag, New York.

\item M\"{u}ller P, Quintana F (2004). \textquotedblleft Nonparametric
Bayesian Data Analysis.\textquotedblright\ \textit{Statistical Science}, 
\textit{19}, 95--110.

\item M\"{u}ller P, Rosner G, Iorio MD, MacEachern S (2005).
\textquotedblleft A Nonparametric Bayesian Model for Inference in Related
Studies.\textquotedblright\ \textit{Applied Statistics}, \textit{54},
611--626.

\item Neal R (2003). \textquotedblleft Slice Sampling (with
Discussion).\textquotedblright\ \textit{Annals of Statistics}, \textit{31},
705--767.

\item Nychka D (2000). \textquotedblleft Spatial-Process Estimates as
Smoothers.\textquotedblright\ In \textit{Smoothing and Regression:
Approaches, Computation, and Application}, pp. 393--424. John Wiley and
Sons, New York.

\item O'Hagan A, Forster J (2004). \textit{Kendall's Advanced Theory of
Statistics: Bayesian Inference, volume 2B}. Arnold, London.

\item Perman M, Pitman J, Yor M (1992). \textquotedblleft Size-Biased
Sampling of Poisson Point Processes and Excursions.\textquotedblright\ 
\textit{Probability Theory and Related Fields}, \textit{92}, 21--39.

\item Pitman J (1995). \textquotedblleft Exchangeable and Partially
Exchangeable Random Partitions.\textquotedblright\ \textit{Probability
Theory and Related Fields}, \textit{102}, 145--158.

\item Pitman J (1996). \textquotedblleft Some Developments of the
Blackwell-MacQueen Urn Scheme.\textquotedblright\ In T Ferguson, L Shapeley,
J MacQueen (eds.), \textit{Statistics, Probability and Game Theory}. Papers
in Honor of David Blackwell, pp. 245--268. Institute of Mathematical
Sciences, Hayward, CA.

\item Pitman J, Yor M (1997). \textquotedblleft The Two-Parameter
Poisson-Dirichlet Distribution Derived from a Stable
Subordinator.\textquotedblright\ \textit{Annals of Probability}, \textit{25}%
, 855--900.

\item Plummer M, Best N, Cowles K, Vines K (2006). \textquotedblleft CODA:
Convergence Diagnosis and Output Analysis for MCMC.\textquotedblright\ 
\textit{R News}, \textit{6}, 7--11.

\item Prado R, West M (2010). \textit{Time Series: Modeling, Computation,
and Inference}. Chapman and Hall/CRC.

\item Quintana F, Iglesias P (2003). \textquotedblleft Bayesian Clustering
and Product Partition Models.\textquotedblright\ \textit{Journal of the
Royal Statistical Society}: \textit{Series B}, \textit{65}, 557--574.

\item R Development Core Team (2015). \textit{R: A Language and Environment
for Statistical Computing}. R Foundation for Statistical Computing, Vienna,
Austria.

\item Rasmussen C, Ghahramani Z (2002). \textquotedblleft Infinite of
Mixtures of Gaussian Process Experts.\textquotedblright\ In T Diettrich, S
Becker, Z Ghahramani (eds.), \textit{Advances in Neural Information
Processing Systems}, \textit{volume 14}, pp. 881--888. The MIT Press,
Cambridge, MA.

\item Regazzini E, Lijoi A, Pr\"{u}nster I (2003). \textquotedblleft
Distributional Results for Means of Normalized Random Measures with
Independent Increments.\textquotedblright\ \textit{Annals of Statistics},%
\textit{\ 31}, 560--585.

\item Robert C, Casella G (2004). \textit{Monte Carlo Statistical Methods
(2nd Ed.)}. Springer, New York.

\item Rosenbaum P, Rubin D (1983a). \textquotedblleft The Central Role of
the Propensity Score in Observational Studies for Causal
Effects.\textquotedblright\ \textit{Biometrika}, \textit{70}, 41--55.

\item Rosenbaum P, Rubin D (1984). \textquotedblleft Reducing Bias in
Observational Studies Using Subclassification on the Propensity
Score.\textquotedblright\ \textit{Journal of the American Statistical
Association}, \textit{79}, 516--524.

\item Rosenthal R (1994). \textquotedblleft Parametric Measures of Effect
Size.\textquotedblright\ In H Cooper, L Hedges (eds.), \textit{The Handbook
of Research Synthesis}, pp. 231--244. Russell Sage Foundation, New York.

\item Schafer J, Kang J (2008). \textquotedblleft Average Causal Effects
from Nonrandomized Studies: A Practical Guide and Simulated
Example.\textquotedblright\ \textit{Psychological Methods}, \textit{13},
279--313.

\item Schervish M (1995). \textit{Theory of Statistics}. Springer-Verlag,
New York.

\item Scott D (1992). \textit{Multivariate Density Estimation: Theory,
Practice, and Visualization}. John Wiley and Sons, New York.

\item Sethuraman J (1994). \textquotedblleft A Constructive Definition of
Dirichlet Priors.\textquotedblright\ \textit{Statistica Sinica}, \textit{4},
639--650.

\item Sethuraman J, Tiwari R (1982). \textquotedblleft Convergence of
Dirichlet Measures and the Interpretation of Their
Parameters.\textquotedblright\ In S Gupta, J Berger (eds.), \textit{%
Statistical Decision Theory and Related Topics III: Proceedings of the Third
Purdue Symposium}, Volume, pp. 305--315. Academic Press, New York.

\item Silverman B (1986). \textit{Density Estimation for Statistics and Data
Analysis}. Chapman and Hall, Boca Raton, Florida.

\item Stroud J, M\"{u}ller P, Sans\'{o} B (2001). \textquotedblleft Dynamic
Models for Spatiotemporal Data.\textquotedblright\ \textit{Journal of the
Royal Statistical Society}, \textit{Series B}, \textit{63}, 673--689.

\item Stuart E (2010). \textquotedblleft Matching Methods for Causal
Inference: A Review and a Look Forward.\textquotedblright\ \textit{%
Statistical Science}, \textit{25}, 1--21.

\item Thistlewaite D, Campbell D (1960). \textquotedblleft
Regression-Discontinuity Analysis: An Alternative to the Ex-post Facto
Experiment.\textquotedblright\ \textit{Journal of Educational Psychology}, 
\textit{51}, 309--317.

\item Thompson S, Sharp S (1999). \textquotedblleft Explaining Heterogeneity
in Meta-Analysis: A Comparison of Methods.\textquotedblright\ \textit{%
Statistics in Medicine}, \textit{18}, 2693--2708.

\item van der Linden W (2015). \textit{Handbook of Modern Item Response
Theory (Second Edition)}. CTB McGraw-Hill, Monterey, CA.

\item Verbeke G, Molenbergs G (2000). \textit{Linear Mixed Models for
Longitudinal Data}. Springer-Verlag, New York.

\item Walker S, Damien P, Laud P, Smith A (1999). \textquotedblleft Bayesian
Nonparametric Inference for Random Distributions and Related
Functions.\textquotedblright\ \textit{Journal of the Royal Statistical
Society}, \textit{Series B}, \textit{61}, 485--527.

\item Wilk M, Gnanadesikan R (1968). \textquotedblleft Probability Plotting
Methods for the Analysis of Data.\textquotedblright\ \textit{Biometrika}, 
\textit{55}, 1--17.

\item Wu TF, Lin CJ, Weng R (2004). \textquotedblleft Probability Estimates
for Multi-Class Classification by Pairwise Coupling.\textquotedblright\ 
\textit{Journal of Machine Learning Research}, \textit{5}, 975--1005.
\end{description}

\newpage

\begin{center}
{\Large Appendix A:\ Technical Preliminaries}
\end{center}

We use the following notation for random variables and probability measures
(functions) (e.g., Berger \&\ Casella, 2002\nocite{CasellaBerger02};
Schervish, 1995\nocite{Schervish95}). A random variable is denoted by an
italicized capital letter, such as $Y$, with $Y$ a function from a sample
space $\mathcal{Y}$ to a set of real numbers. A realization of that random
variable is denoted by a lower case, with $Y=y$. Also, $F(\cdot )$ (or $%
P(\cdot )$) is the probability measure (function) that satisfies the
Kolmogorov probability axioms, having sample space domain $\mathcal{Y}$ and
range $(0,1)$. Then $F(B)$ (or $P(B)$) denotes the probability of any event $%
B\subset \mathcal{Y}$ of interest. Throughout, a probability measure is
denoted by a capital letter such as $F$ (or $G$ or $\Pi $ resp., for
example), and $f(y)$ ($g(y)$ and $\pi (y)$, resp.) is the corresponding
probability density of $y$, if $Y$ is discrete or continuous.

If $Y$ is a continuous random variable, with sample space $\mathcal{Y}%
\subset \mathbb{R}^{d}$ ($d\in \mathbb{Z}_{+}$) and Borel $\sigma $-field $%
\mathcal{B}(\mathcal{Y})$, and with a probability density function (p.d.f.) $%
f$ defined on $\mathcal{Y}$ such that $\tint f(y)\mathrm{d}y=1$, then the
probability measure of $f$ is given by $F(B)=\tint_{B}f(y)\mathrm{d}y$ for
all $B\in \mathcal{B}(\mathcal{Y})$, with $f(y)=\mathrm{d}F(y)/\mathrm{d}%
y\geq 0$. If $Y$ is a discrete random variable with a countable sample space 
$\mathcal{Y}=\{y_{1},y_{2},\ldots \}$ and Borel $\sigma $-field $\mathcal{B}(%
\mathcal{Y})$, and with probability mass function (p.m.f.) $f$ defined on $%
\mathcal{Y}$ such that $\tsum\nolimits_{k=1}^{\infty }f(y_{k})=1$, then the
probability measure of $f$ is given by $F(B)=\tsum\nolimits_{\{k:y_{k}\in
B\}}f(y_{k})$ for all $B\in \mathcal{B}(\mathcal{Y})$, with $0\leq
f(y)=P(Y=y)\leq 1\mathrm{.}$ Also, a cumulative distribution function
(c.d.f.) is the probability measure $F(y):=F(B)=P(Y\leq y)$, where $%
B=\{y^{\prime }:-\infty <y^{\prime }\leq y\}$. The c.d.f. $F(y)$ corresponds
to a p.d.f. $f(y)=\mathrm{d}F(y)/\mathrm{d}y$ if $Y$ is continuous; and
corresponds to a p.m.f. $f(y)=P(Y=y)$ if $Y$ is discrete. A multidimensional
random variable, $\boldsymbol{Y}=(Y_{1},\ldots ,Y_{d})$, has c.d.f. $%
F(y_{1},\ldots ,y_{d})=P(Y_{1}\leq y_{1},\ldots ,Y_{k}\leq y_{d})$.

$Y\sim F$ means that the random variable $Y$ has a distribution defined by
the probability measure $F$. Thus $F$ is also called a distribution.
Notation such a $Y\sim F(y\,|\,x)$ (or $Y\sim F(y\,|\,\mathbf{x})$, resp.)
means that the random variable $Y\,|\,x$ (the random variable, $Y\,|\,%
\mathbf{x}$, resp.) has a distribution defined by a probability measure $%
F(\cdot \,|\,x)$ conditionally on the value $x$ of a variable $X$ (or has a
distribution $F(\cdot \,|\,\mathbf{x})$ conditionally on the value $\mathbf{x%
}=(x_{1},\ldots ,x_{p})$ of $p$ variables). Sometimes a probability
distribution (measure) is notated without the $\cdot \,|$ symbol.

We denote \textrm{N}$(\cdot \,|\,\mu ,\sigma ^{2})$, $\Phi (\cdot )=$ 
\textrm{N}$(\cdot \,|\,0,1)$, \textrm{Ga}$(\cdot \,|\,a,b)$, \textrm{IG}$%
(\cdot \,|\,a,b)$, \textrm{U}$(\cdot \,|\,a,b)$, \textrm{Be}$(\cdot
\,|\,a,b) $, \textrm{GIG}$(\cdot \,|\,a,b,q)$, \textrm{N}$(\cdot \,|\,%
\boldsymbol{\mu },\boldsymbol{\Sigma })$, \textrm{IW}$(\cdot \,|\,d,%
\boldsymbol{S})$, and \textrm{Di}$(\cdot \,|\,\boldsymbol{\alpha })$, as the
probability measures of a normal distribution with mean and variance
parameters $(\mu ,\sigma ^{2})$; the standard normal distribution; the gamma
distribution with shape and rate parameters $(a,b)$; the inverse-gamma
distribution with shape and rate parameters $(a,b)$; the uniform
distribution with minimum and maximum parameters $(a,b)$; the beta
distribution with shape parameters $(a,b)$; the inverse-Gaussian
distribution with mean $\mu >0$ and shape $\lambda >0$; the generalized
inverse-Gaussian distribution with parameters $(a,b,q)\in \mathbb{R}%
_{+}\times \mathbb{R}_{+}\times \mathbb{R}$; the multivariate normal
distribution for a random vector $(Y_{1},\ldots ,Y_{d})$, with mean $%
\boldsymbol{\mu }=(\mu _{1},\ldots ,\mu _{d})^{\intercal }$ and $d\times d$
variance-covariance matrix $\boldsymbol{\Sigma }$; the inverted-Wishart
distribution with degrees of freedom $d$ and scale matrix $\boldsymbol{S}$;
and of the Dirichlet distribution with precision parameters $\boldsymbol{%
\alpha }$, respectively. These distributions are continuous and define
p.d.f.s\ denoted by lower-case letters, with \textrm{n}$(\cdot \,|\,\mu
,\sigma ^{2})$, \textrm{n}$(\cdot \,|\,0,1)$, \textrm{ga}$(\cdot \,|\,a,b)$, 
\textrm{ig}$(\cdot \,|\,a,b)$, \textrm{u}$(\cdot \,|\,a,b)$, \textrm{be}$%
(\cdot \,|\,a,b)$, \textrm{gig}$(\cdot \,|\,a,b,q)$, \textrm{n}$(\cdot \,|\,%
\boldsymbol{\mu },\boldsymbol{\Sigma })$, \textrm{iw}$(\cdot \,|\,d,%
\boldsymbol{S})$, and \textrm{di}$(\cdot \,|\,\boldsymbol{\alpha })$,
respectively. The p.d.f. equations are found in statistical distribution
textbooks (e.g., Johnson et al., 1994\nocite{JohnsonKotzBalkrishnan94}).

\bigskip \newpage

\begin{center}
{\Large Appendix B:\ BNP Data Analysis Exercises}
\end{center}

You may use the Bayesian regression software to answer the following data
analysis exercises. Relevant data sets are available from the software, and
are described under the Help menu.

\begin{enumerate}
\item \textbf{(Regression) \ }Perform regression analyses of the
PIRLS100.csv data from the 2006 Progress in International Reading Literacy
Study (PIRLS). Data are from a random sample of 100 students who each
attended one of 21 U.S. low income schools. The dependent variable, zREAD,
is a continuous-valued student literacy score. Also, consider eight
covariates\ (predictors):\ a binary (0,1) indicator of male status (MALE),
student age (AGE), size of student's class (CLSIZE), percent of English
language learners in the student's classroom (ELL), years of experience of
the student's teacher (TEXP), years of education of the student's teacher
(EDLEVEL), the number of students enrolled in the student's school (ENROL),
and a 3-point rating of the student's school safety (SCHSAFE = 1 is highest;
SCHSAFE = 3 is lowest). There is also data on a school ID\ (SCHID) grouping
variable.\newline
Analyze the data set using the regression models listed below, after
performing a z-score transformation of the 8 covariates using the menu
option: Modify Data Set \TEXTsymbol{>}\ Simple variable transformations 
\TEXTsymbol{>}\ Z-score. The SCHID grouping variable may be included in a
multi-level regression model.\newline
-- \ ANOVA/linear DDP\ model (Dirichlet process (DP)\ mixture of linear
regressions).\newline
-- \ An infinite-probits model.\newline
-- \ Another Bayesian nonparametric (infinite) mixture of regression models.%
\newline
(a)\ Now, describe (represent)\ each of the models, mathematically and in
words.\newline
(b) Summarize the results of your data analyses, while focusing on the
relevant results. Describe the relationship between the dependent variable
and each covariate. For the infinite-probits model, describe the clustering
behavior in the posterior predictive distribution (density)\ of the
dependent variable, conditionally on values of one or two covariates of your
choice. For the DDP\ model, graphically describe the clustering behavior of
the random intercept and slope parameters.\newline
(c)\ Evaluate how well the data support the assumptions of other parametric
models, such as the linearity of regression, the normality of the regression
errors, and the normality of the random intercept and slope coefficients.
Fit a normal linear regression and normal random-effects (multi-level HLM)\
models to the data, and compare the predictive fit between the linear
model(s) and the Bayesian nonparametric models above.

\item \textbf{(Binary Regression) \ }Analyze the data using the following
binary regression models, with binary (0,1)\ dependent variable READPASS,
and the 8 z-scored covariates.\newline
-- \ ANOVA/linear DDP\ logit (or probit) model (a DP mixture of regressions).%
\newline
-- \ An infinite-probits model for binary regression.\newline
-- \ Another Bayesian nonparametric (infinite) mixture of regression models.%
\newline
(a)\ Describe (represent)\ each of the models, mathematically and in words.%
\newline
(b) Summarize the results of your data analyses, while focusing on the
relevant results. Describe the relationship between the dependent variable
and each covariate. For the infinite-probits model, describe the clustering
behavior in the posterior predictive distribution (density)\ of the
(latent)\ dependent variable, conditionally on values of 1 or 2 covariates
of your choice. For the DDP\ model, graphically describe the clustering
behavior of the random intercept and slope parameters.\newline
(c)\ Evaluate how well the data support assumptions of other parametric
models, namely, the unimodality of the (latent)\ regression errors, and the
normality of the random intercept and slope coefficients. This will require
fitting a probit (or logit)\ linear and/or random-effects regression model
to the data. Compare the fit between the probit (or logit)\ model(s) and the
Bayesian nonparametric models listed above.

\item (\textbf{Causal Analysis}) \ Analyze the PIRLS100.csv data using a
Bayesian nonparametric regression model.\ Investigate the causal effect of
large (versus small) class size (\textbf{CLSIZE}) on reading performance (%
\textbf{zREAD}), in the context of a regression discontinuity design (e.g.,
Cook, 2008\nocite{Cook08}). For the data set, use a Modify Data Set menu
option to construct a new (treatment)\ variable defined by a (0,1) indicator
of large class size, named \textbf{LARGE}, where \textbf{LARGE }= 1 if 
\textbf{CLSIZE}\ $\geq $ 21, and zero otherwise. Perform a regression of 
\textbf{zREAD} on the predictors (\textbf{LARGE}, \textbf{CLSIZE}), in order
to infer the causal effect of large class size (\textbf{LARGE }= 1) versus
small class size (\textbf{LARGE }= 0), on the variable \textbf{zREAD},
conditionally on \textbf{CLSIZE }= 21. Do so by inferring the coefficient
estimates of the covariate \textbf{LARGE}, and by performing posterior
predictive inferences of \textbf{zREAD}, conditionally on \textbf{LARGE }= 1
versus \textbf{LARGE }= 0. Under relatively mild conditions, such
comparisons provide inferences of causal effects (of large versus small
class size), as if the treatments were randomly assigned to subjects
associated with class sizes in a small neighborhood around \textbf{CLSIZE }=
21. A key condition (assumption)\ is that students have imperfect control
over the \textbf{CLSIZE }variable, around the value of \textbf{CLSIZE }= 21
(Lee, 2008\nocite{Lee08}; Lee\ \&\ Lemieux, 2010\nocite{LeeLemieux10}).

\item (\textbf{3-level data)} \ Analyze the Classroom data set
(classroom300.csv) using a Bayesian nonparametric regression model. In the
data, students (Level 1) are nested within classrooms (Level 2) nested
within schools (Level 3). A classroom identifier (classid) provides a
level-2 grouping variable. A school identifier (schoolid) provides a level-3
grouping variable. Model mathgain as the dependent variable, defined by
student gain in mathematical knowledge. Also, for the model, include at
least three covariates: student socioeconomic status (ses), the level of
experience of the student's teacher (yearsteaching), and student house
poverty level (housepov).

\item \textbf{(Longitudinal Data Analysis) } Fit a Bayesian nonparametric
regression model to analyze the GPA.csv data. Investigate the relationship
between gpa and the covariates of time, job type, gender status, and
possibly student ID. Consider performing an auto-regressive time-series
analysis of the data, using the model. For this, you can use a Modify Data
Set menu option to construct lag-terms of the dependent variable (for
selected lag order) and then include them as additional covariates in the
regression model.

\item \textbf{(Meta Analysis)} \ Fit a Bayesian nonparametric regression
model to perform a meta-analysis of the Calendar.csv data. The dependent
variable is \textbf{Effect} size, here defined by the unbiased standardized
difference (Hedges, 1981\nocite{Hedges81}) between mean student achievement
in schools that follow a year-round calendar, and (minus)\ the mean student
achievement in schools that follow the traditional nine-month calendar. The
covariates are standard error of the effect size (\textbf{SE\_ES}), and
study publication \textbf{Year}; and the observation weights are given by
the variable \textbf{weight} (= 1/\textbf{Var}). The overall effect-size,
over studies, is represented by the model's intercept parameter. In your
report of the data analysis results, common on the impact of publication
bias on estimates of the overall effect size. Publication bias may be
assessed by inferring the slope coefficient estimate of the \textbf{SE\_ES}
covariate; and/or by posterior predictive inferences of the Effect dependent
variable, conditionally over a range of values of \textbf{SE\_ES}. Such
inferences provide regression analyses for a funnel plot (Thompson \&\
Sharp, 1999\nocite{ThompsonSharp99}).

\item \textbf{(Survival Analysis of censored data) \ }Use a Bayesian
nonparametric regression model to perform a survival analysis of the
larynx.csv data or the bcdeter.csv data. The larynx data set has dependent
variable \textbf{logyears}; and covariates of patient \textbf{age}, cancer 
\textbf{stage}, and diagnosis year (\textbf{diagyr}). The observations of 
\textbf{logyears} are either right-censored or uncensored, as indicated by
the variables \textbf{LB} and \textbf{UB}. The bcdeter data set has
dependent variable\textbf{\ logmonths}, and treatment type indicator
covariates (\textbf{radioth}, \textbf{radChemo}). The \textbf{logmonths }%
observations are either right-censored, interval-censored, or uncensored, as
indicated by the variables \textbf{logLBplus1} and \textbf{logUBplus1}.

\item (\textbf{Item Response Analysis}) Use a BNP model to perform an item
response (IRT)\ analysis of either the NAEP75.csv data set, or the
Teacher49.csv data set. The NAEP75 data has binary item-response scores (0 =
Incorrect, 1 = Correct). The Teacher49 data has ordinal item response scores
(0, 1, or 2). Each data set has information about an examinee, per data row,
with examinee item responses in multiple data columns. For either data set,
use a "vectorize"\ Modify Data Set menu option, to collapse the multiple
item response variables (columns)\ into a single column dependent variable,
and to construct item dummy (0 or $-$1) covariates. Report the (marginal)\
posterior estimates of examinee ability, item difficulty (for each test
item), as well as the estimates of the other model parameters.
\end{enumerate}
\end{subequations}

\end{document}